\documentclass[techreport]{acmart}

\usepackage{booktabs} 

\usepackage{mathtools}
\usepackage{amsmath,amssymb,amsfonts}
\usepackage{graphicx}
\usepackage{subfig}
\usepackage{url}
\usepackage{todonotes}
\usepackage[noend]{algpseudocode}
\usepackage{algorithm}
\usepackage{xspace}
\usepackage{enumitem}
\usepackage{makecell}

\usepackage{sidecap}
\sidecaptionvpos{figure}{c}

\def\sortidu{\textsc{sortidu}\xspace}
\def\reorder{\textsc{reorder}\xspace}
\def\scircuit{\textsc{shortc}\xspace}
\def\gpu{\textsc{GPU-Join}\xspace}

\def\ego{\textsc{SuperEGO}\xspace}
\def\lss{\textsc{LSS}\xspace}

\def\cooc{\textit{CoocTexture}\xspace}
\def\colorhist{\textit{ColorHist}\xspace}
\def\layout{\textit{LayoutHist}\xspace}
\def\susy{\textit{SuSy}\xspace}
\def\songs{\textit{Songs}\xspace}

\def\datasetsyn{\textit{Syn-}\xspace}

\def\dsynthah{\textit{Syn16D2M}\xspace} 
\def\dsynthai{\textit{Syn32D2M}\xspace} 
\def\dsynthaj{\textit{Syn64D2M}\xspace}

\usepackage{xpatch}

\makeatletter
\xpatchcmd{\ps@firstpagestyle}{Manuscript submitted to ACM}{}{\typeout{First patch succeeded}}{\typeout{first patch failed}}
\xpatchcmd{\ps@standardpagestyle}{Manuscript submitted to ACM}{}{\typeout{Second patch succeeded}}{\typeout{Second patch failed}}    \@ACM@manuscriptfalse
\makeatother

\renewcommand\footnotetextcopyrightpermission[1]{} 

\title{GPU Accelerated Similarity Self-Join for Multi-Dimensional Data}

\author{Michael Gowanlock}
\orcid{0000-0002-0826-6204}
\affiliation{%
  \institution{School of Informatics, Computing, \& Cyber Systems\\ Northern Arizona University}
  \streetaddress{PO Box 5693}
  \city{Flagstaff}
  \state{AZ}
  \postcode{86011}
}
\email{michael.gowanlock@nau.edu}

\author{Ben Karsin}
\affiliation{%
  \institution{University of Hawai`i at M\=anoa\\Department of Information and Computer Sciences}
  \streetaddress{POST Building, Rm 317, 1680 East-West Road}
  \city{Honolulu}
  \state{HI}
  \postcode{96822}
}
\email{karsin@hawaii.edu}

\begin{document}

\begin{abstract}
The self-join finds all objects in a dataset that are within a search distance, $\epsilon$, of each other; therefore, the self-join is a building block of many algorithms. We advance a GPU-accelerated self-join algorithm targeted towards high dimensional data. The massive parallelism afforded by the GPU and high aggregate memory bandwidth makes the architecture well-suited for data-intensive workloads. 
We leverage a grid-based, GPU-tailored index to perform range queries.   We propose the following optimizations:  $(i)$ a trade-off between candidate set filtering and index search overhead by exploiting properties of the index; $(ii)$ reordering the data based on variance in each dimension to improve the filtering power of the index; and $(iii)$ a pruning method for reducing the number of expensive distance calculations. Across most scenarios on real-world and synthetic datasets, our algorithm outperforms the parallel state-of-the-art approach. 
Exascale systems are converging on heterogeneous distributed-memory architectures.  We show that an entity partitioning method can be utilized to achieve a balanced workload, and thus good scalability for multi-GPU or distributed-memory self-joins. 
\end{abstract}

\maketitle

%
%


\section{Introduction}\label{sec:intro}
A fundamental database operation and an intensely studied problem in database research is the similarity self-join. Self-similarity joins are often part of existing algorithms~\cite{ester1996density,koperski1995discovery,Bayardo:2007:SUP:1242572.1242591,Ankerst:1999:OOP:304182.304187,agrawal1993efficient} and are fundamental to many established methods~\cite{Bohm:2000:HPC:354756.354832,bohm2009}. Since the self-join is commonly employed, improving self-join performance has implications for many domains.  The problem is as follows: given a dataset of objects, find all objects that have common attributes based on a similarity metric. This paper focuses on the distance similarity self-join that finds all points that are within the Euclidean distance $\epsilon$ of each other.

Research on the self-join typically addresses either low~\cite{bryan2008compact} or high~\cite{andoni2006near,lieberman2008fast,Dittrich:2001:GSS:502512.502524} dimensionality. In low dimensionality, the data points (or feature vectors) are often more frequently co-located; therefore, there are typically more neighbors on average in comparison to high dimensionality~\cite{Jacox:2008:MSS:1366102.1366104}. Therefore, in low dimensionality, filtering the candidate set of points becomes a performance bottleneck. To reduce the number of comparisons between data points when filtering the candidate set, indexes are used. However, in high dimensionality, index searches become increasingly exhaustive due to the well-known curse of dimensionality~\cite{bellman1961,Durrant:2009:NNM:1558391.1558528,Volnyansky:2009:CDP:1637863.1638181,Kim2015}, where the index needs to search a large fraction of the database, potentially degrading the search to the performance level obtained by a brute force search.  Thus, searches in high dimensionality can be prohibitive.  In this work, we focus on optimizing the high dimensionality self-join.

There have been two recent trends in computer architecture that have been helping to guide the direction of many research communities.  First, modern graphics processing units (GPUs) are frequently used to attain high computational throughput through
massive parallelism and high memory bandwidth on a range of problems in the constituent fields of computer science. Second, distributed-memory systems that are composed of hundreds or thousands of heterogeneous nodes are becoming more commonplace, where nodes are comprised of one or more accelerators.  This recent development has been motivated by the high performance computing community striving to reach exascale. In this context, GPU-efficient algorithms are essential to achieving peak performance on GPUs and heterogeneous systems.
Thus, many large-scale data analytics applications will rely on GPU-efficient algorithms, including the distance similarity self-join for high dimensional data --- the subject of this work.  This paper makes the following novel contributions:
\begin{itemize}[leftmargin=*]
\setlength{\itemsep}{1pt}
\setlength{\parskip}{0pt}
\setlength{\parsep}{0pt}  
\item Leveraging an efficient indexing scheme for the GPU, we exploit the trade-off between index filtering power and search cost to improve the overall performance of searching high dimensional feature spaces.
\item  We improve the filtering power of the index by reordering the data in each dimension using statistical properties of the data distribution. We show that this is particularly important when exploiting the trade-off outlined above. 
\item We mitigate the performance cost of reducing index filtering power by proposing a technique that prunes the candidate set by comparing points based on an un-indexed dimension.    
\item We show that on the worst-case data distribution for our approach, we achieve significantly better performance than the state-of-the-art on the same scenario. This suggests that the performance of the GPU-accelerated self-join is resilient to the data distribution, making the approach well-suited for many application scenarios.
\item We evaluate our approach on 5 real-world and 3 synthetic datasets and show that our GPU accelerated self-join outperforms the state-of-the-art parallel algorithm in the literature.
\item The self-join is an expensive operation. We show initial insights into the scalability of the self-join on multi-GPU and distributed-memory systems, and demonstrate that an entity partitioning strategy can be used to achieve good load balancing.
\end{itemize}

The paper is outlined as follows: Section~\ref{sec:background} provides background material, 
Section~\ref{sec:probstatement} formalizes the problem and discusses previous work that we employ, Section~\ref{sec:optimizations} presents the novel methods we use to improve high dimensional self-join performance, Section~\ref{sec:expereval} illustrates our performance results, Section~\ref{sec:scalability_pathways} discusses the scalability of the self-join,  and we conclude the work in Section~\ref{sec:conclusions}.

\section{Background}\label{sec:background}

The self-join is conceptualized as a join relationship between a database table and itself. Works that optimize joining two different tables are relevant to the self-join. Also, efficient indexing structures for performing neighborhood range queries are relevant, as indexes reduce the number of point comparisons. We also review the distributed-memory self-join. 


\subsection{Similarity-Joins and the State-of-the-art}
The similarity-join is a well-studied algorithm~\cite{Bohm:2000:HPC:354756.354832,Bohm2001,Dittrich:2001:GSS:502512.502524,Arasu:2006:EES:1182635.1164206,Bayardo:2007:SUP:1242572.1242591,kalashnikov2013}. Here, we discuss those works that address high-dimensional data. GESS~\cite{Dittrich:2001:GSS:502512.502524} assigns feature vectors to hypercubes, and then performs an intersection query on these hypercubes to compute the similarity join. The method relies on data replication and duplication removal from the result set. LSS~\cite{lieberman2008fast} utilizes the GPU, and transforms the similarity join into a sort-and-search problem. Interval searches are needed, and the authors use space filling curves to reduce interval size and search overhead. The Super-EGO algorithm~\cite{kalashnikov2013} has been shown to be effective for similarity-joins on both low and high-dimensional data. The algorithm uses the ``epsilon grid order''~\cite{bohm2001epsilon} method. It uses a non-materialized grid to find nearby points that may be within the search distance. Then, based on a query point's cell and nearby cells, the algorithm prunes the search for nearby points by filtering cells by their $n$-dimensional coordinates.  However, unlike previous work~\cite{bohm2001epsilon}, Super-EGO exploits the properties of the data to improve performance. The authors find that Super-EGO~\cite{kalashnikov2013} outperforms both GESS~\cite{Dittrich:2001:GSS:502512.502524}, and LSS~\cite{lieberman2008fast}, so we compare our work to Super-EGO.           

\subsection{GPU Self-Join on Low-Dimensional Data}
Gowanlock and Karsin~\cite{GowanlockKarsin2018} studied the self-join on the GPU for low-dimensional data using a grid-based index, and show that between 2 and 6 dimensions, the self-join outperforms both canonical search-and-refine and state-of-the-art approaches (i.e., Super-EGO). However, they show that index search overhead increases exponentially with dimensionality, and they limit their work to low dimensional data.  In this work, we implement methods that utilize a similar indexing structure, but we utilize optimizations for high-dimensional self-joins.

\subsection{Indexing on the GPU}
We utilize the GPU due to its high memory bandwidth and computational throughput. There are two major indexing strategies for the GPU: $(i)$ index-trees, similar to those that have been shown to provide good performance on the CPU, such as the R-tree~\cite{Guttman-R_tree}; or $(ii)$ non-hierarchical indexes, such as grids or binning.  
Several works propose efficient indexes for points or other objects on the GPU~\cite{bohm2009,Zhang:2012:USH:2390226.2390229,Kim2015,Gowanlock2015,Gowanlock2016,KIM2018195}. 

Kim et al.~\cite{Kim2015} designed an R-tree for the GPU to optimize index searches that avoids many of the drawbacks of executing tree traversals on the GPU.  Later, the same research group presented a hybrid approach~\cite{KIM2018195} that splits the R-tree between the CPU and GPU by assigning parts of the algorithm with more regular and irregular instruction flows to the GPU and CPU, respectively.  The approach is effective because irregular instruction flows reduce parallel efficiency on the GPU.
This work suggests that, while an indexing scheme with less filtering power results in more distance comparisons, a more regularized instruction flow may lead to better performance on the GPU. Thus, it may be more effective to have an indexing scheme that is less work-efficient (i.e., on average there is a larger candidate set of points to filter), instead of a more work-efficient index that has a smaller candidate set.

\subsection{Distributed-Memory Implementations}
High-dimensional self-joins are expensive for even moderate dataset sizes. 
While this work and the most closely related works described above focus on scaling \emph{up} 
the self-join,
there are several works that consider scaling \emph{out} by using multiple compute nodes in a cluster.
A MapReduce~\cite{dean2008mapreduce} implementation is presented by Fries et al.~\cite{Fries2014}. The benefit of using MapReduce is that the user does not need to implement many of the technical details related to concurrency.  Fries et al.~\cite{Fries2014} propose a self-join for high-dimensional data using MapReduce, where they show that they can significantly reduce data duplication by using a method they term ``dimension groups'', where they perform the self-join on subsets of the data dimensions first, and then union these subsets to obtain the final result. They show that their approach yields much lower data duplication than previous work~\cite{seidl2013mr}.

A major issue in any distributed-memory implementation (MapReduce or otherwise) is that the data needs to be distributed to the compute nodes, and this may require duplicating a fraction of the data points on the nodes. While the focus of this work is a single-GPU self-join for high-dimensional data, we demonstrate that we can obtain no data duplication using an entity-based partitioning strategy using an MPI implementation, and that the additional communication overhead associated with our approach is negligible in comparison to the computation of the self-join as performed at each node. 


\section{Problem Statement and Previous Insights}\label{sec:probstatement}
\subsection{Problem Statement}
Let $D$ be a database of points (or feature vectors). Each point in the database is denoted as $p_i$, where $i=1,\dots,|D|$. Each $p_i\in D$ has coordinates in $n$-dimensions, where each coordinate is denoted as $x_j$ where $j=1,\dots,n$, and $n$ is the number of dimensions of the point/feature vector. Thus, the coordinates of point $p_i$ are denoted as: $p_i=(x_1,x_2,\dots,x_{n})$. We refer to the $x_j$-coordinate value of point $p_i$ as $p_i(x_j)$.
As with prior related work (Section~\ref{sec:background}), we use the Euclidean distance (the $L^2$ norm). We find all pairs of points that are within $\epsilon$ distance of each other.  We say that points $a \in D$ and $b \in D$ are within $\epsilon$ distance when $dist(a,b)\leq\epsilon$, where $dist(a,b)=\sqrt{\sum_{j=1}^n(a(x_j)-b(x_j))^2}$.  Thus, the result of the self-join are tuples ($a\in D$, $b\in D$), where $a$ and $b$ are within $\epsilon$ distance of each other. 

The self-join is a special case of the similarity-join. If we let $E$ be a set of entry points in an index (defined similarly to the definition above) and $Q$ be a set of query points, the similarity join finds all points in $Q$ within $\epsilon$ distance of $E$, i.e., $Q \bowtie_\epsilon E$. In contrast, the self-join is simply $E \bowtie_\epsilon E$. Thus, the self-join is relevant to the similarity-join problem as well.

By comparing all points to each other, the worst-case complexity is $O(|D|^2)$, which can be simply implemented as a nested loop join~\cite{Jacox:2007:SJT:1206049.1206056}.  However, as mentioned above, indexes can be used to reduce the number of comparisons between points.     

We perform the self-join entirely in-memory. The dataset, result set, and working memory do not exceed main memory on the host. However, the global memory of the GPU is limited, and ensuring that the self-join does not exceed global memory capacity would make it intractable for many self-join scenarios. Therefore, while $D$ does not exceed global memory capacity, we allow the result set to exceed global memory capacity. The result set is typically much larger than $D$, and this enables a larger number of self-join scenarios (larger result sets, or databases).

\subsection{Leveraging Previous Insights}\label{sec:previous_work}
In Section~\ref{sec:optimizations} we outline our novel methods for performing the self-join in high dimensions. However, we leverage several optimizations from the literature that are relevant to the self-join. In particular, we use the grid-based GPU index presented by Gowanlock \& Karsin~\cite{GowanlockKarsin2018}, that builds on prior work~\cite{Gowanlock2017}. These papers also advanced a batching scheme, which we use to process self-join result sets that may exceed the GPU's global memory capacity. We briefly describe the batching and indexing techniques that we reuse from the literature, and note that we cannot directly use the low-dimensional methods~\cite{GowanlockKarsin2018} for high-dimensional self-joins.     

\subsubsection{Grid-Based Indexing on the GPU}\label{sec:grid_index}
We utilize an $n$-dimensional grid index for computing the self-join.  As mentioned in Section~\ref{sec:background}, the state-of-the-art join algorithm for high dimensional data, Super-EGO~\cite{kalashnikov2013}, also uses a grid-based technique for efficiently computing the self-join. We refer the reader to the work of Gowanlock et al.~\cite{Gowanlock2017} for an in-depth description of the index, which the authors used in 2-D for clustering with DBSCAN~\cite{ester1996density}. A major difference between the indexing scheme used in this work and that of Gowanlock et al.~\cite{Gowanlock2017} is that we do not index empty cells, as the space complexity would be intractable for high dimensions (as also discussed by Gowanlock and Karsin~\cite{GowanlockKarsin2018}).

The GPU grid index from Gowanlock et al.~\cite{Gowanlock2017} is constructed as follows. On the host, the data points, $D$, are sorted into unit-length bins in each dimension. This ensures that data points nearby each other in the $n$-dimensional space are nearby each other in memory. Each grid cell is of length $\epsilon$, which ensures that for a given point, only the adjacent cells need to be searched to find points that are within the $\epsilon$ distance. This bounds the search on the GPU to regularize the instruction flow. 
For demonstrative purposes and without loss of generality, we assume a grid with edges starting at 0 in each dimension, and assign points to cells by simply computing the cell's $n$-dimensional coordinates from the point's ($p_i$) coordinates as follows: ($x_1/\epsilon$, $x_2/\epsilon$, $\dots$ , $x_n/\epsilon$). The points are not stored within the grid structure, rather, the points belonging to a grid cell are stored in a lookup array that each grid cell references when finding the points contained within. This minimizes the memory needed to store the points within a grid cell. Lastly, since we only store non-empty cells, we create a lookup array that stores the linearized ids of the non-empty grid cells. 
As shown by Gowanlock and Karsin~\cite{GowanlockKarsin2018}, the storage requirements simplify to the size of the dataset, $O(|D|)$. This compact index structure allows more space on the GPU to be allocated for larger input and result sets.

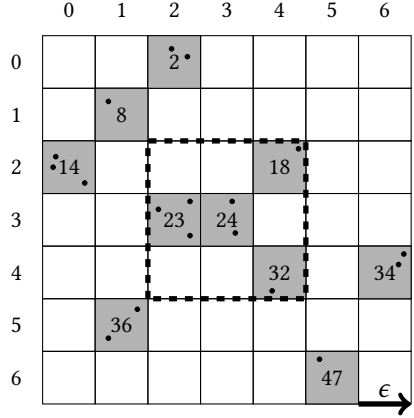
\begin{figure}[t]
\centering
    

\begin{tikzpicture}

\def\scale{0.7}

\foreach \x in {0,1,2,3,4,5,6} {
	\foreach \y in {0,1,2,3,4,5,6} {
		\draw (\x*\scale,\y*\scale) rectangle (\x*\scale+1*\scale,\y*\scale+1*\scale);
	}
}

\foreach \x in {6,5,4,3,2,1,0}{
\node at (\x*\scale+0.5*\scale,7.5*\scale) {\x};
}

\foreach \x in {0,1,2,3,4,5,6}{
\node at (-0.5*\scale,6*\scale-\x*\scale+0.5*\scale) {\x};
}


\foreach \x/\y/\z in {0/4/14,1/1/36,2/3/23,1/5/8,5/0/47,3/3/24,4/4/18,4/2/32,6/2/34,2/6/2} {
	\draw [fill=white!70!black] (\x*\scale,\y*\scale) rectangle (\x*\scale+1*\scale,\y*\scale+1*\scale);
	\node at (\x*\scale+0.5*\scale,\y*\scale+0.5*\scale) {{\z}};
}

\foreach \x/\y in {1.8/1.8, 1.25/1.25, 0.8/4.2, 0.2/4.5, 0.25/4.7, 1.25/5.75, 2.45/6.75, 2.75/6.6, 2.2/3.7, 2.8/3.2, 2.8/3.85, 3.6/3.85, 3.66/3.25, 4.36/2.15, 4.86/4.85, 6.86/2.85, 6.76/2.65, 5.26/0.85  } {
	\draw [fill=black] (\x*\scale,\y*\scale) circle (0.03cm);
}


\draw [thick,dashed, line width=2pt] (2*\scale,2*\scale) rectangle (4*\scale+1*\scale,4*\scale+1*\scale);

\draw [->,line width=2.0pt, scale=1] (6*\scale,0*\scale) -- (7*\scale,0*\scale);

\node at (6.5*\scale,0.25*\scale) {{\Large$\epsilon$}};


\end{tikzpicture}     
    \caption{Example of searching the grid index in 2-D. The non-empty cells are shaded. Numbers refer to linearized cell ids.}
   \label{fig:GPU_index}
\end{figure}

Figure~\ref{fig:GPU_index} shows an example grid in 2-D. The non-empty gray cells with linearized cell ids are shown. Consider a point in cell 24. To find all of its neighbors within $\epsilon$, it needs to search the adjacent cells (and its origin cell), which are encompassed by the black dashed line. In $n$ dimensions, there are $3^n$ cells to search. However, the points in the cells are not guaranteed to be within $\epsilon$; therefore, distance calculations between the query point and all of the points in the cells bounded by the dashed outline (cells 18, 23, 24, 32 in the figure) are needed to determine which ones are within the search distance $\epsilon$.       

The self-join is executed on the GPU with a \emph{kernel} that uses $|D|$ threads. Each thread is assigned a point and finds all neighbors within the $\epsilon$ distance. The threads write the result to a buffer as key/value pairs, where the key is a point and the value if a point within $\epsilon$ of the key. After all threads have completed finding their respective neighbors within $\epsilon$, the key/value pairs are sorted on the GPU, and returned to the host.

Bounding the search to neighboring cells using the grid reduces thread divergence, 
which is known to degrade GPU performance~\cite{PascalPerformance}. Also, the grid structure improves data locality among threads within a warp, increasing \emph{coalesced} memory accesses and unified cache usage as global memory loads are stored in the L1 cache on NVIDIA Maxwell and Pascal architectures~\cite{PascalPerformance}.

\subsubsection{Batching: Enabling Large Result Sets}\label{sec:batching}   
A drawback of the GPU is that it has limited global memory. The self-join result sets can be much larger than the input dataset size $|D|$. Thus, an efficient batching scheme is needed to incrementally compute the entire self-join result. However, this is typically a bigger problem in low-dimensionality, as more points are co-located, due to smaller (hyper)volumes. 

To ensure that the self-join result set does not exceed global memory capacity, we employ the method from Gowanlock et al.~\cite{Gowanlock2017}, and provide a summary of their work as follows. First, a kernel is executed that finds all of the neighbors within $\epsilon$ for a fraction of the points in the dataset, which is used to estimate the size of the total result set. This kernel invocation takes negligible time in comparison to the total time needed to execute the self-join, as only a fraction of points are searched, and the kernel only returns an integer (the number of points within $\epsilon$, not the actual result as key/value pairs).  The number of batches, denoted as $n_b$, is computed based on a batch size, denoted $b_s$, and the estimated total result set size.  
Even if the result set for a given $\epsilon$ would not overflow the global memory capacity of the GPU, the batching scheme is still used as it allows for overlap of data transfers to and from the GPU, computation on the GPU itself, and host-side operations. It is preferable to overlap these components of the algorithm to maximize concurrent use of resources. In all of our experiments, we use a minimum of 3 CUDA streams, and hence batches, i.e., $n_b\geq3$. We allocate 3 \emph{pinned memory} buffers on the host, as it is needed for overlapping data transfers to and from the GPU, and has a higher data transfer rate~\cite{PascalPerformance}. For result set sizes that exceed $3\times10^8$, we set a batch size of $b_s=10^8$ (the total neighbors found within $\epsilon$ of each point). Thus, each stream is assigned a buffer of size $b_s=10^8$.

\section{High-Dimensional Self-Join Optimizations}\label{sec:optimizations}
As described in Section~\ref{sec:intro} the performance of algorithms that compute the distance similarity self-join is largely dependent on the dimensionality of the input dataset. This section outlines self-join optimizations tailored for high-dimensional datasets, particularly in the context of the grid-based indexing scheme presented in Section~\ref{sec:grid_index}.

\subsection{Index Selectivity to Target High Dimensionality}\label{sec:indexkdim}
In high dimensionality, there are fewer co-located neighbors because, as the hypervolume increases, the distance between objects increases~\cite{Jacox:2008:MSS:1366102.1366104}.  However, with increasing dimension, index filtering power decreases and search performance degrades. There is a trade-off between index filtering power and search overhead. The drawback is that an index with less filtering power will yield larger candidate set sizes that need to be evaluated by computing the Euclidean distance between a query point and all candidate points. The benefit, however, is that the search required to find the candidate set will incur less overhead. 

The GPU is a suitable architecture for making a trade-off between less filtering power and more search overhead. This is because the GPU is designed to achieve high computational throughput and thus excels at tasks like computing the distances between points in parallel. Thus, to avoid the overheads associated with searches in higher dimensions, we provide a trade-off using a less rigorous index search for increased filtering overhead. To illustrate why this trade-off is important, in the context of the grid indexing scheme, the number of adjacent cells required to check is $3^n$; in 2-D, this is only 9 cells, but in 6-D, this is 729 cells.
We decrease the filtering power and search overhead by indexing only $k$ dimensions of the $n$-dimensional points, where $2\leq k<n$, thus projecting the points into $k$ dimensions. To resolve whether points are within $\epsilon$ of the query point, we compute the Euclidean distance in all $n$ dimensions, and thus obtain the correct result.  Since we index in fewer than $n$ dimensions, each cell has $n-k$ unconstrained dimensions, resulting in less filtering power.  The decrease in the number of adjacent cells searched is denoted as $l(n,k)=(3^n-3^k)/3^n$. For example, if 5-D points are indexed in 3-D, then the loss is: $l(n,k)=(3^5-3^3$)/$3^5$=88.9\%, and the number of searched cells decreases by that percentage (assuming the data is uniformly distributed in each dimension). 





\subsection{Dimensionality Reordering by Variance}\label{sec:reorder}
Index searches are increasingly exhaustive and more expensive in higher dimensions. The statistical properties of high dimensional feature vectors can be exploited to improve the filtering power of the index to prune the search space and eliminate points that are not within $\epsilon$ (e.g., see \cite{kalashnikov2013} in related work, Section~\ref{sec:background}). The dimensions of the data that have the greatest variance should improve the pruning power of index searches.  Since we may not index all dimensions, as discussed in Section~\ref{sec:indexkdim}, it is important to select dimensions that optimize the ability of the index to prune the search and minimize the size of the candidate set. Otherwise, if we simply select the first $k$ dimensions in the dataset, we may inadvertently select dimensions to index that yield a minimal degree of pruning power.   

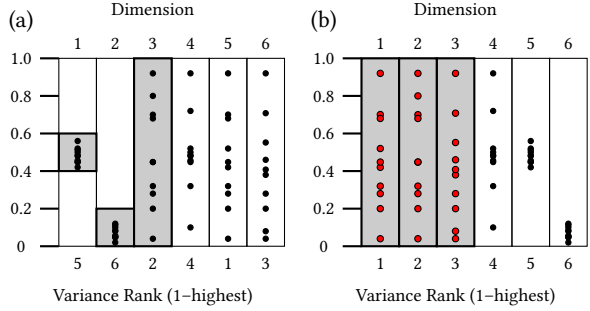
\begin{figure}[tp]
\centering
    \subfloat{

\begin{tikzpicture}[scale=1.0]

\def\scale{0.9}

\node at (0,7.5) {(a)};

\def\xscale{0.5}
\foreach \x in {1,2,3,4,5,6} {
		\draw (\x*\xscale,5*\scale) rectangle (\x*\xscale+1*\xscale,14*\xscale);
}

\def\xscale{0.5}
\foreach \x\y\z in {1/11/12, 2/9/10, 3/9/14} {
		\draw [color=black, fill=white!50!black!40!, thick](\x*\xscale,\y*\xscale) rectangle (\x*\xscale+1*\xscale,\z*\xscale);
}

\foreach \x in {6,5,4,3,2,1}{
\node at (\x*\xscale+0.5*\xscale,8*\scale) {\footnotesize{\x}};
}

\foreach \x\y in {6/3,5/1,4/4,3/2,2/6,1/5}{
\node at (\x*\xscale+0.5*\xscale,4.75*\scale) {\footnotesize{\y}};
}

\node at (3*\xscale+0.5*\xscale,4.25*\scale) {\footnotesize{Variance Rank (1--highest)}};

\node at (3*\xscale+0.5*\xscale,8.5*\scale) {\footnotesize{Dimension}};

\def\yscale{0.5}
\foreach \y\z in {9/0,10/0.2,11/0.4,12/0.6,13/0.8,14/1.0}
    {      
      \draw [thick](0.25,\y*\yscale) -- (0.5,\y*\yscale);
      \node at (0,\y*\yscale) {\footnotesize{\z}};
    }

\foreach \x\y in {0.75/1.2, 0.75/1.25, 0.75/1.3, 0.75/1.28, 0.75/1.12, 0.75/1.21, 0.75/1.05, 0.75/1.12, 0.75/1.4,0.75/1.14 } {
    \filldraw (1*\x,\y+5*\scale) circle (1pt);
}

\foreach \x\y in {1.25/0.25, 1.25/0.2, 1.25/0.3, 1.25/0.12, 1.25/0.28, 1.25/0.05, 1.25/0.21, 1.25/0.14, 1.25/0.12 ,1.25/0.13} {
    \filldraw (1*\x,\y+5*\scale) circle (1pt);
}

\foreach \x\y in {1.75/1.75, 1.75/2.0, 1.75/2.3, 1.75/1.12, 1.75/0.5, 1.75/0.8, 1.75/0.1, 1.75/0.70, 1.75/1.12 ,1.75/1.7} {
    \filldraw (1*\x,\y+5*\scale) circle (1pt);
}

\foreach \x\y in {2.25/2.3, 2.25/1.8, 2.25/1.12, 2.25/0.8, 2.25/0.25, 2.25/1.2, 2.25/1.21, 2.25/1.14, 2.25/1.3 ,2.25/1.25} {
    \filldraw (1*\x,\y+5*\scale) circle (1pt);
}

\foreach \x\y in {2.75/1.05, 2.75/1.752, 2.75/0.8, 2.75/1.7, 2.75/1.3, 2.75/0.70, 2.75/2.3, 2.75/0.1, 2.75/1.12, 2.75/0.5} {
    \filldraw (1*\x,\y+5*\scale) circle (1pt);
}

\foreach \x\y in {3.25/0.95, 3.25/1.02, 3.25/1.15, 3.25/1.77, 3.25/0.2, 3.25/1.38, 3.25/2.3, 3.25/0.1, 3.25/0.5, 3.25/0.70} {
    \filldraw (1*\x,\y+5*\scale) circle (1pt);
}

\end{tikzpicture}     
        }
        \subfloat{


\begin{tikzpicture}[scale=1.0]

\def\scale{0.9}

\node at (0,7.5) {(b)};

\def\xscale{0.5}
\foreach \x in {1,2,3,4,5,6} {
		\draw (\x*\xscale,5*\scale) rectangle (\x*\xscale+1*\xscale,14*\xscale);
}

\foreach \x in {6,5,4,3,2,1}{
\node at (\x*\xscale+0.5*\xscale,8*\scale) {\footnotesize{\x}};
}

\node at (3*\xscale+0.5*\xscale,8.5*\scale) {\footnotesize{Dimension}};

\foreach \x\y in {6/6,5/5,4/4,3/3,2/2,1/1}{
\node at (\x*\xscale+0.5*\xscale,4.75*\scale) {\footnotesize{\y}};
}

\node at (3*\xscale+0.5*\xscale,4.25*\scale) {\footnotesize{Variance Rank (1--highest)}};

\def\yscale{0.5}
\foreach \y\z in {9/0,10/0.2,11/0.4,12/0.6,13/0.8,14/1.0}
    {        
      \draw [thick](0.25,\y*\yscale) -- (0.5,\y*\yscale);
      \node at (0,\y*\yscale) {\footnotesize{\z}};
    }

\def\xscale{0.5}
\foreach \x\y\z in {1/9/14, 3/9/14, 2/9/14} {
		\draw [color=black, fill=white!50!black!40!, thick](\x*\xscale,\y*\xscale) rectangle (\x*\xscale+1*\xscale,\z*\xscale);
}

\foreach \x\y in {2.75/1.2, 2.75/1.25, 2.75/1.3, 2.75/1.28, 2.75/1.12, 2.75/1.21, 2.75/1.05, 2.75/1.12, 2.75/1.4,2.75/1.14 } {
    \filldraw (1*\x,\y+5*\scale) circle (1pt);
}

\foreach \x\y in {3.25/0.25, 3.25/0.2, 3.25/0.3, 3.25/0.12, 3.25/0.28, 3.25/0.05, 3.25/0.21, 3.25/0.14, 3.25/0.12 ,3.25/0.13} {
    \filldraw (1*\x,\y+5*\scale) circle (1pt);
}


\foreach \x\y in {1.25/1.75, 1.25/2.0, 1.25/2.3, 1.25/1.12, 1.25/0.5, 1.25/0.8, 1.25/0.1, 1.25/0.70, 1.25/1.12 ,1.25/1.7} {
    \filldraw [color=red, draw=black] (1*\x,\y+5*\scale) circle (1.2pt);
}

\foreach \x\y in {2.25/2.3, 2.25/1.8, 2.25/1.12, 2.25/0.8, 2.25/0.25, 2.25/1.2, 2.25/1.21, 2.25/1.14, 2.25/1.3 ,2.25/1.25} {
    \filldraw (1*\x,\y+5*\scale) circle (1pt);
}

\foreach \x\y in {0.75/1.05, 0.75/1.752, 0.75/0.8, 0.75/1.7, 0.75/1.3, 0.75/0.70, 0.75/2.3, 0.75/0.1, 0.75/1.12, 0.75/0.5} {
    \filldraw [color=red, draw=black] (1*\x,\y+5*\scale) circle (1.2pt);
}

\foreach \x\y in {1.75/0.95, 1.75/1.02, 1.75/1.15, 1.75/1.77, 1.75/0.2, 1.75/1.38, 1.75/2.3, 1.75/0.1, 1.75/0.5, 1.75/0.70} {
    \filldraw  [color=red, draw=black] (1*\x,\y+5*\scale) circle (1.2pt);
}











\end{tikzpicture}     
        }
    \caption{Dimensionality reordering by variance on a 6-D dataset having $|D|=10$ indexing $k=3$ dimensions. (a) initial input dataset; (b) reordering the point coordinates from largest to smallest variance in each dimension. Red points denote those used to index $k=3$ dimensions based on high variance in corresponding dimensions. Shaded cells denote indexed area.}
   \label{fig:selectivity}
\end{figure}


Figure~\ref{fig:selectivity}~(a) shows an example dataset of 10 points in 6 dimensions generated in the range [0,1]. 
We can see that the first two dimensions have a low degree of variance.  Thus, if we index $k=3$ dimensions (and not all $n=6$), we will have a low amount of index filtering power due to the low variance of the first two dimensions.
If we assume that the grid cells are of length $\epsilon=0.2$, we find that dimensions 1 and 2 will only produce a single cell in their dimensions (denoted by the shaded regions). Thus, when searching the index for points, dimensions 1 and 2 do not reduce the number of points that may be within $\epsilon$. However, we can select the dimensions with greater variance to improve filtering power.  For instance, dimensions 5, 3 and 6 in Figure~\ref{fig:selectivity}~(a) have the greatest variance.  If we reorder the data by decreasing variance, then we obtain Figure~\ref{fig:selectivity}~(b). 
Now, each of the first 3 dimensions spans 5 grid cells (assuming $\epsilon=0.2$). Consequently, when the index is searched, it will return cells with \emph{fewer} points on average in comparison to the original input distribution in Figure~\ref{fig:selectivity}~(a). We note that in Figure~\ref{fig:selectivity}, it seems like the \emph{number} of cells should be maximized and not the \emph{variance}. While data with high variance will tend to produce more cells, it is possible to have many cells in a dimension with low variance (e.g., one point per cell, and the remaining points in a single cell, as in dimension 4 in Figure~\ref{fig:selectivity}~(b)).

To re-order the dimensions by their variance, we use a sample of 1\% of $|D|$ and estimate the variance in each
dimension.
Then, we reorder the coordinate values in each dimension of $p_i\in D$, such that the values are in descending order from highest to lowest variance. Thus, when we index the first $k$ dimensions (Section~\ref{sec:indexkdim}), they potentially have greater filtering power than the initial input dataset. Reordering dimensions does not impact the correctness of the result, as we are simply swapping the coordinate values of the points. This requires $O(|D|n)$ work, which is negligible when compared to performing the self-join. We denote reordering the data by variance in each dimension as \reorder. 
If we index $k<n$ dimensions, but do not use \reorder, we simply index the first $k$ dimensions of the input dataset.

\subsection{Searching on an Un-indexed Dimension}
\begin{figure}[t]
  \centering
  \includegraphics[width=0.45\textwidth]{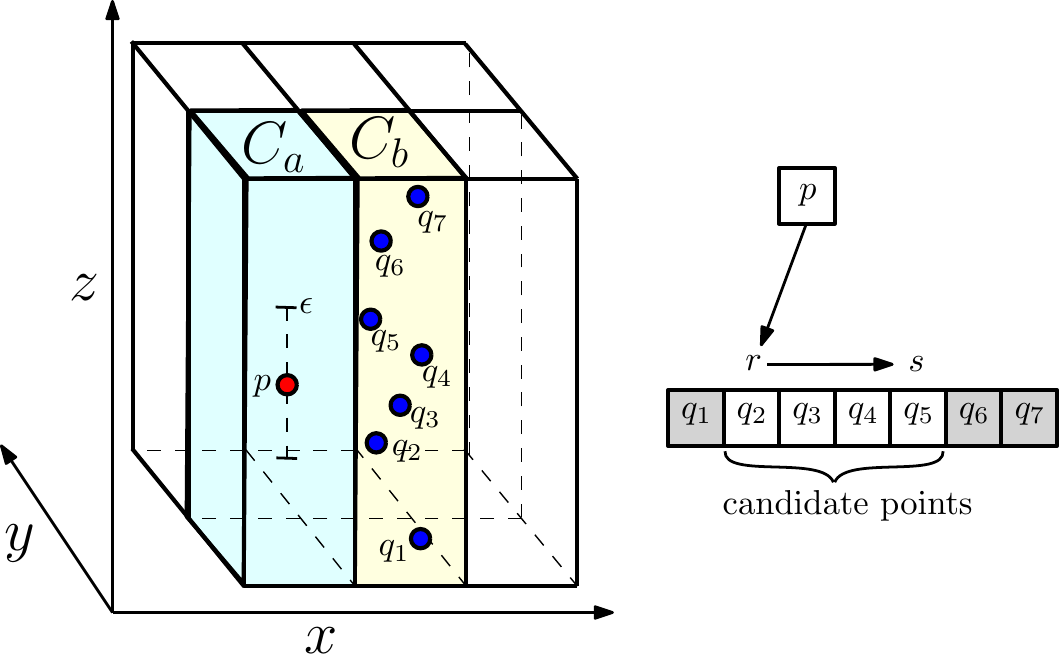}
  \caption{Example of the \sortidu optimization.  We first sort points within each cell by $z$-coordinate.  We then search query point $p$ 
into the list of $C_b$ to find $r$ and scan and compare points until $s$ is reached.}
  \label{fig:sortIndex}
\end{figure}

By indexing only $k<n$ dimensions, we reduce the indexing overhead by reducing the number of grid cells, which is exponential with $k$.  However, this
comes at the cost of reduced filtering power, resulting in more distance calculations.  In this section we introduce a technique of searching on an un-indexed
dimension to further reduce the number of necessary distance calculations.  Consider an input set with $n$ dimensions that is indexed on $k<n$ dimensions
using the indexing scheme presented in Section~\ref{sec:grid_index}.  For a given point $p$ in cell $C_a$ and neighbor cell $C_b$, we  compare $p$ 
to each point $q \in C_b$ to determine if $p$ and $q$ are within distance $\epsilon$ of each other.  Since we have indexed $k$ dimensions, the points contained
in $C_b$ are only filtered by these $k$ dimensions.  Thus, if we consider dimension $u$ that is \emph{not} indexed, each point in $C_b$ can have any value
in this dimension.  Currently, we must perform a distance comparison on all $q \in C_b$, which includes such points that may be very distant from $p$
in dimension $u$ (i.e., $|p(u) - q(u)| > \epsilon$).  Therefore, we propose an optimization called \sortidu to only compare $p$ 
with $q \in C_b$ if they are within 
$\epsilon$ distance along the $u$-coordinate.  We accomplish this by first sorting the points within each cell by increasing $u$-coordinate.  
When comparing $p$ with all $q \in C_b$, we first search $p(u)$ into the points in $C_b$ to find the point $r$ with the smallest $u$-coordinate
that is still within $\epsilon$ of $p$ (i.e, $|p(u) - r(u)| \le \epsilon$).  We then scan points in $C_b$ by increasing $u$ coordinate until we reach
point $s$ with more than the $\epsilon$ distance in the $u$-coordinate (i.e., $|p(u) - s(u)| > \epsilon$).  Figure~\ref{fig:sortIndex} illustrates an example
of the \sortidu optimization.  In this example, the $z$-axis is not indexed and we use the \sortidu to reduce the number
of candidate points we have to consider, from $q_1, q_2,\dots,q_7$ (7 points) to $q_2,\dots,q_5$ (4 points). We note that we only perform this optimization on 
one un-indexed dimension and all other un-indexed dimensions remain unfiltered.

If every point in $C_b$ is within $\epsilon$ from $p$, the \sortidu optimization provides no performance improvement.
However, for reasonably small $\epsilon$ values, this can significantly reduce the number of candidate points.  This comes at the cost of sorting and
searching.  If we consider that cell $C_b$ has $|C_b|$ points, we must perform $|C_b|$ distance calculations without the \sortidu optimization.
However, \sortidu reduces this to $(\log{|C_b|} + m)$ calculations, where $m$ is the number of points in $C_b$ with $u$-coordinate within
$\epsilon$ of $p(u)$.  While \sortidu requires that we sort points within each cell, we only have to sort once for all point 
evaluations.
We can apply the \sortidu optimization even when we index all dimensions (i.e., $k=n$).  In this
case, we sort each cell by one of the indexed dimensions; however, we expect more significant benefits
when we apply \sortidu to an un-indexed dimension.

\subsection{Short Circuiting the Distance Calculation}\label{sec:scircuit}
The number of floating point operations needed to compute the Euclidean distance between two $n$-dimensional points is $3n$. Thus, the distance calculation cost increases with dimensionality. 
Since the calculation of Euclidean distance is based on the sum of values, we can incrementally compute it.
During computation, if the partial sum exceeds $\epsilon$ before the entire Euclidean distance is calculated, we
know the two points are not within $\epsilon$ distance of each other and can stop further calculation.
This type of \emph{sort-circuiting} is a well-known optimization and has been used in other works such as \ego~\cite{kalashnikov2013}, and we show that it leads to significant performance gains in some scenarios.  We denote this optimization
as \scircuit.

\subsection{Outline of the Algorithm}\label{sec:algorithm_outline}

Algorithm~\ref{alg:self_join} outlines \gpu, which begins by re-ordering the input set, $D$, by variance, if \reorder is enabled
(line~\ref{algline.reorder}).  
Next, the index is computed using the dataset and the number of indexed dimensions, $k$ (line~\ref{algline.index}). Then, the number of batches, $n_b$, to be executed are computed using the batch size, $b_s$
(line~\ref{algline.numbatches}).
The algorithm then loops over all of the batches (line~\ref{algline.loopbatches}) and executes them on the GPU (line~\ref{algline.kernel}, detailed below).  
The result of each batch is stored as key/value pairs, where the key is the a query point and the value is a point within $\epsilon$ distance (i.e., for each query point $p_i$ there may be multiple resulting pairs $(p_i, p_j)$). Since the keys are often redundant (multiple points are within $\epsilon$ of a given point), they are stored without redundant information using constructNeighborTable, and stored into the final result (line~\ref{algline.result}).

Each batch is executed by running {\textsc{SelfJoinKernel}\xspace}, a GPU kernel. First, the result set for the batch is initialized (line~\ref{algline.resultsetinit}), the global id of the thread is obtained (line~\ref{algline.globalid}), and the point, $p_i \in D$ is computed as a function of the global id (line~\ref{algline.getpoint}). Next, all of the adjacent non-empty cells are computed from $G$ (the index), $k$, and the point (line~\ref{algline.getadjcells}). The algorithm loops over each neighbor cell (line~\ref{algline.loopcells}), and computes the distance between the query point and all of the points within the neighbor cell to determine if they are within $\epsilon$ (line~\ref{algline.calcdist}). After all neighbor cells have been processed, the points within $\epsilon$ of the query point assigned to the thread are added to the result set, and the kernel returns (lines~\ref{algline.resultset}--\ref{algline.return}). The execution of the calcDistancePts function differs if \sortidu or \scircuit are enabled.

\begin{algorithm}
\caption{\gpu Algorithm}
\label{alg:self_join}
\begin{algorithmic}[1]

\begin{footnotesize}
\Procedure{\gpu}{$D$, $\epsilon$, $n$, $k$, $b_s$}
\State $D \leftarrow$ importData()\label{algline.importdata}
\State $D \leftarrow$ reorderVariance($D$)\label{algline.reorder}
\State $G \leftarrow$ constructIndex($D$, $k$)\label{algline.index}
\State $n_b \leftarrow$ computeNumBatches($b_s$)\label{algline.numbatches}
\State result $\leftarrow \emptyset$\label{algline.resultinit}
\For {$i \in$  1,$\dots$,$n_b$} \label{algline.loopbatches}
\State kernelResult[i] $\leftarrow$ selfJoinKernel($D$, $G$, $n$, $k$, $\epsilon$)\label{algline.kernel}
\State result $\leftarrow$ result $\cup$ constructNeighborTable(kernelResult[i])\label{algline.result}
\EndFor

\State \Return
\EndProcedure

\item[]

\Procedure{SelfJoinKernel}{$D$, $G$, $n$, $k$, $\epsilon$}
\State resultSet $\leftarrow \emptyset$\label{algline.resultsetinit}
\State gid $\leftarrow$ getGlobalId()\label{algline.globalid}
\State point $\leftarrow$ getPoint(gid, $D$)\label{algline.getpoint}
\State adjCells $\leftarrow$ getAdjCells($G$, $k$, point)\label{algline.getadjcells}
\For {cell $\in$ adjCells.min,$\dots$,adjCells.max }\label{algline.loopcells}
\State pntResult $\leftarrow$ pntResult $\cup$ calcDistancePts(point, cell, $n$, $\epsilon$)\label{algline.calcdist}
\EndFor
\State resultSet $\leftarrow$ resultSet $\cup$ pntResult\label{algline.resultset}

\State \Return resultSet\label{algline.return}
\EndProcedure

\end{footnotesize}
\end{algorithmic}
\end{algorithm}

\begin{figure}[t]
\centering

\begin{tikzpicture}[scale=0.61]

  \def\xscale{1}
  \def\yscale{1}
  \def\length{12}

  \foreach \x in {0,...,6}{
  \draw [-] (0+\xscale,0.5*\x) -- (\xscale*\length+1,0.5*\x);
  }

  \foreach \x in {1,...,5}{
  \node (batchid) at (\xscale-0.25,0.5*\x-0.25) {\small{\x}};
  }
  \node (batchid) at (\xscale-0.25,0.5*6-0.25) {\small{$\dots$}};

  \foreach \x in {1,...,9} {
  	\draw [solid] (1.5*\x-0.5,0) -- (1.5*\x-0.5,3);
  }

  \foreach \x in {1,...,5} {
    \node (HtoD)  at (\x*1.5+0.25,-0.25+\x*0.5) {\small{HtoD}};
    \node (GPU)   at (\x*1.5+1.75,-0.25+\x*0.5) {\small{Kernel}};
    \node (DtoH)  at (\x*1.5+3.25,-0.25+\x*0.5) {\small{DtoH}};
    \node (Table) at (\x*1.5+4.75,-0.25+\x*0.5) {\small{Table}};
  }

   \draw [->, ultra thick] (0+\xscale,-0.5) -- (\xscale*\length+1,-0.5);

   \node (time) at (7,-0.9) {Time}; 
   \node (Batches)[rotate=90] at (-0.1,1.5) {Batch Num.};








\end{tikzpicture}     
    \caption{Illustrative example of pipelined operations in \gpu. \textit{HtoD} denotes a host-to-device data transfer, \textit{Kernel} denotes execution of the GPU kernel, \textit{DtoH} denotes a device-to-host data transfer, and \textit{Table} denotes constructing the neighbor table on the host.}
   \label{fig:pipeline_batches}
\end{figure}
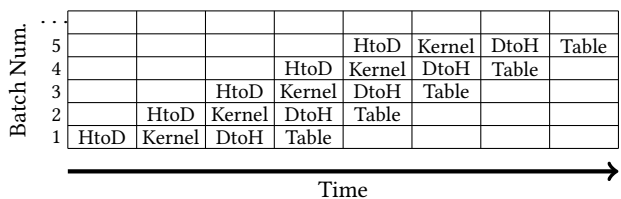

While Algorithm~\ref{alg:self_join} appears to be sequential, we execute several of the algorithm batching scheme tasks concurrently. The loop on line~\ref{algline.loopbatches} in Algorithm~\ref{alg:self_join} is executed in parallel.   Figure~\ref{fig:pipeline_batches} shows an overview of a pipeline of tasks that are executed concurrently, where we use a balanced pipeline for illustrative purposes. For each batch, the kernel parameters relevant to the batch are sent to the GPU (\emph{HtoD}), the GPU computes the self-join result for the batch (\emph{Kernel}), and then the result is transferred back to the host (\emph{DtoH}), and finally the neighbor table is constructed (\emph{Table}). This pipeline shows that up to four concurrent tasks can be executed at the same time. Particularly important is transferring of the result sets back to the host (\emph{DtoH}) while the next batch is executing (\emph{Kernel}), which hides much of the data transfer overhead. Note that this pipeline is for illustrative purposes only. In high dimensionality, the algorithm is compute-bound, so \emph{Kernel} is much longer than the other stages. However, if there are smaller workloads that have short kernel execution phases, then the fraction of GPU-related overheads increase (\emph{HtoD} and \emph{DtoH}). In this case, the pipeline is important to ensure that \gpu remains competitive with CPU-only algorithms on small workloads.

\section{Experimental Evaluation}\label{sec:expereval}
\subsection{Datasets}\label{sec:datasets}
We utilize both real and synthetic datasets to evaluate the performance of our algorithm and optimizations, which we summarize in Table~\ref{tab:datasets} and describe below. We use some of the real-world datasets to evaluate both \lss~\cite{lieberman2008fast} and \ego~\cite{kalashnikov2013}. We normalize all datasets in the range [0,1] to conform to the requirements of \ego. Real-world datasets were obtained from UCI ML repository~\cite{Lichman:2013}.
\begin{itemize}[leftmargin=*]
\setlength{\itemsep}{1pt}
\setlength{\parskip}{0pt}
\setlength{\parsep}{0pt}  
\item Co-occurrence Texture, \cooc -- 16-D image features, and 68,040 points. Used in \lss and \ego.  
\item Color Histogram, \colorhist -- 32-D image features, and 68,040 points. Used in \lss and \ego.  
\item Layout Histogram, \layout -- 32-D image features and 66,616 points. Used in \lss and \ego.  
\item Supersymmetry Particles, \susy 18-D -- kinematic properties of 5 million particles from the Large Hadron Collider. Used for classification~\cite{2014NatCo...5E4308B}.    
\item Song Prediction Dataset, \songs 90-D -- extracted features of songs, with 415,345 points. Used for classification~\cite{Bertin-Mahieux2011}.
\end{itemize}

\begin{table}[b]
\centering
\caption{Dataset, data points, $|D|$, and dimension, $n$.}\label{tab:datasets}
\begin{tabular}{|c|c|c|c|c|c|} \hline
Dataset &$|D|$ & $n$ & Dataset &$|D|$ & $n$\\ \hline
\cooc &68,040&16&\songs &515,345&90\\ \hline
\layout &66,616&32&\dsynthah&$2\times10^6$&16\\ \hline
\colorhist&68,040&32&\dsynthai&$2\times10^6$&32\\\hline
\susy&$5\times10^6$&18&\dsynthaj&$2\times10^6$&64\\\hline
\end{tabular}
\end{table}

We use real-world datasets to evaluate performance of high dimensional self-joins, as synthetic datasets are not representative of real-world data. However, the similarity joins in the literature (and our work) rely on statistical techniques for improving index efficacy. The \emph{worst case} scenario for our algorithm is when there is low variance in each dimension, reducing the impact of dimensionality reordering (Section~\ref{sec:reorder}). To evaluate performance on such worst-case inputs, we generate synthetic datasets with an exponential distribution with $\lambda=40$.
We generate 16, 32, and 64-dimensional synthetic datasets, denoted \datasetsyn, with coordinates in [0,1].


\subsection{Experimental Methodology}
The GPU code is written in CUDA and executed on an NVIDIA GP100 GPU with 16 GiB of global memory. The C/C++ host code is compiled with the GNU compiler (v. 5.4.0) and O3 optimization flag.  The platform that executes all experiments has $2\times$ E5-2620 v4 2.1 GHz CPUs, with each having 8 cores (16 total cores). Our self-join CUDA kernel uses 256 threads per block, and uses 32-bit floats for consistency with \ego. We exclude the time to load the dataset and construct the index. We include the time to execute the self-join, store the result set on the host and other host-side operations, and perform the dimensionality reordering.  
Thus, we include the response time of components used in other works to make a fair comparison between approaches.  

We perform experiments using a series of datasets and $\epsilon$ values such 
that we do not have too many
(e.g., $|D|^2$) or too few (e.g., 0) total results. Thus, the values of $\epsilon$ should represent values that are pragmatically useful to a user of the self-join algorithm. For instance, when using the self-join for data analysis, it is not useful to return a large fraction of the total database, $D$.     We denote the selectivity of the self-join as follows:

\begin{equation}
S_D=\frac{|R|-|D|}{|D|},
\end{equation} 
where $|R|$ is the total result set size. This yields the average number of points within $\epsilon$, excluding a point, $p_a$, finding itself (i.e., a result tuple: ($p_a \in D, p_a \in D$)). We report the selectivity in our plots so that our results can be reproduced by other researchers and to demonstrate that the respective experimental scenario is meaningful. Our experiments cover a range of $\epsilon$ values that include those used by Kalashnikov~\cite{kalashnikov2013} to evaluate \ego. 

\subsection{State-of-the-art Implementation (\ego)}\label{sec:exp_ego}
Super-EGO~\cite{kalashnikov2013} performs fast self-joins on multidimensional data and has been shown to outperform other algorithms on low and high dimensional data. We use a multi-threaded implementation of Super-EGO, using 16 threads on 16 physical cores (the maximum number of cores on our platform). We normalize the datasets in the range [0,1] in each dimension, as needed by the algorithm. We compute the total time using the time to ego-sort and join on 32-bit floats and exclude the other components (e.g., loading the dataset and indexing). We ensure self-join correctness by comparing the results of \gpu and \ego. We are grateful to D. Kalashnikov for making \ego publicly available.

\subsection{Performance on Datasets in the SuperEgo Paper}\label{sec:ego_compare}
As $\epsilon$ increases, the workload of the self-join increases for two reasons: $(i)$ index searches are more exhaustive as more data overlaps the query point being searched; and $(ii)$ larger candidate sets need to be filtered using Euclidean distance  calculations. In this section, we compare the performance of \gpu to \lss and \ego, for $n\geq16$ and varying $\epsilon$, using the same datasets as those used by Kalashnikov~\cite{kalashnikov2013}.
We do not use any optimizations except indexing $k<n$ dimensions (Section~\ref{sec:indexkdim}). We discuss how to select $k$ in Section~\ref{sec:eval_dim_reduction_reorder}.

\begin{figure}[tp]
\centering
\subfloat[\colorhist (32-D)]{
        \includegraphics[width=0.22\textwidth, trim={0.5cm 0 0.5cm 0}]{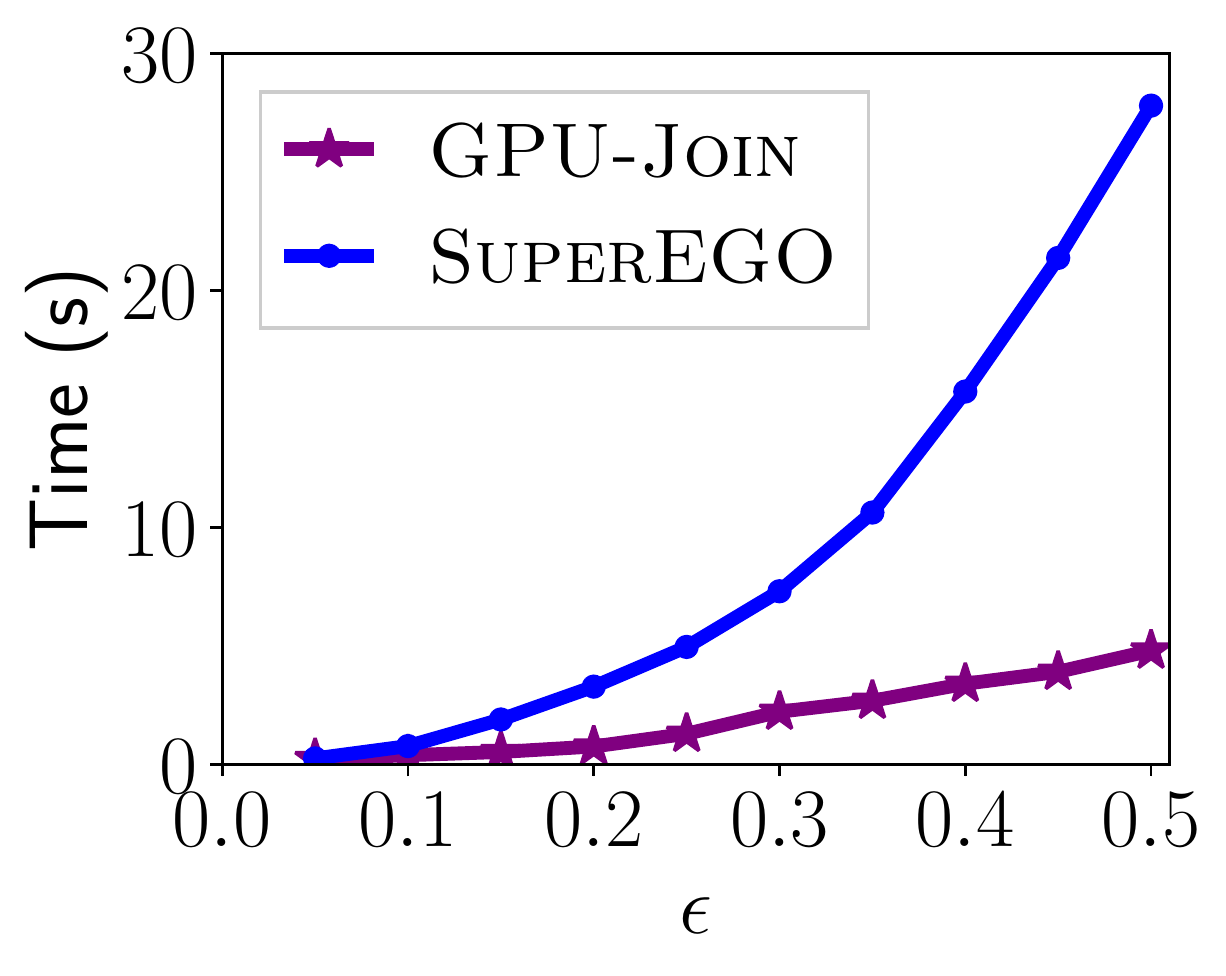}
    }      
\subfloat[\layout (32-D)]{
        \includegraphics[width=0.22\textwidth, trim={0.5cm 0 0.5cm 0}]{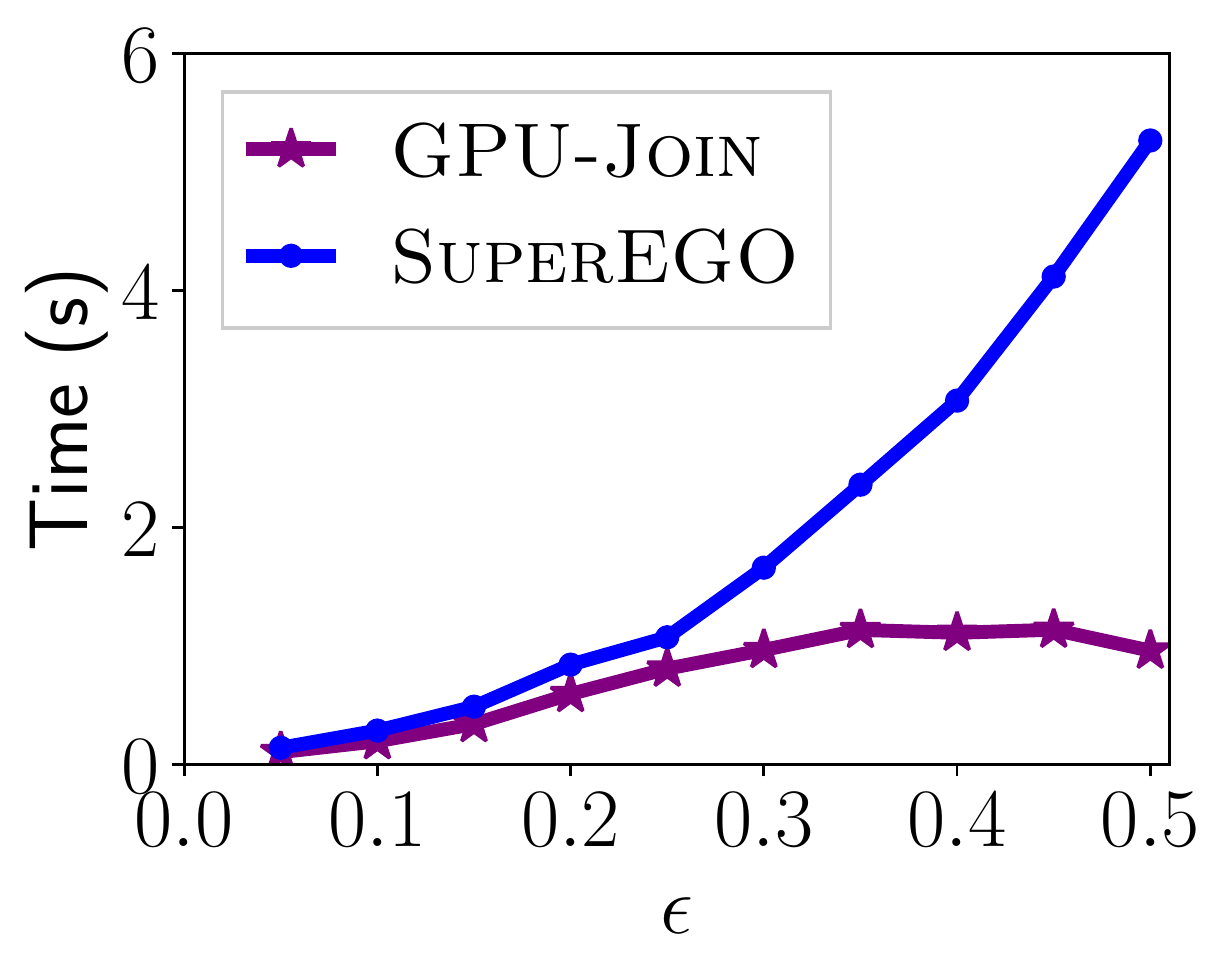}
    } 

\subfloat[\cooc (16-D)]{
        \includegraphics[width=0.22\textwidth, trim={0.5cm 0 0.5cm 0}]{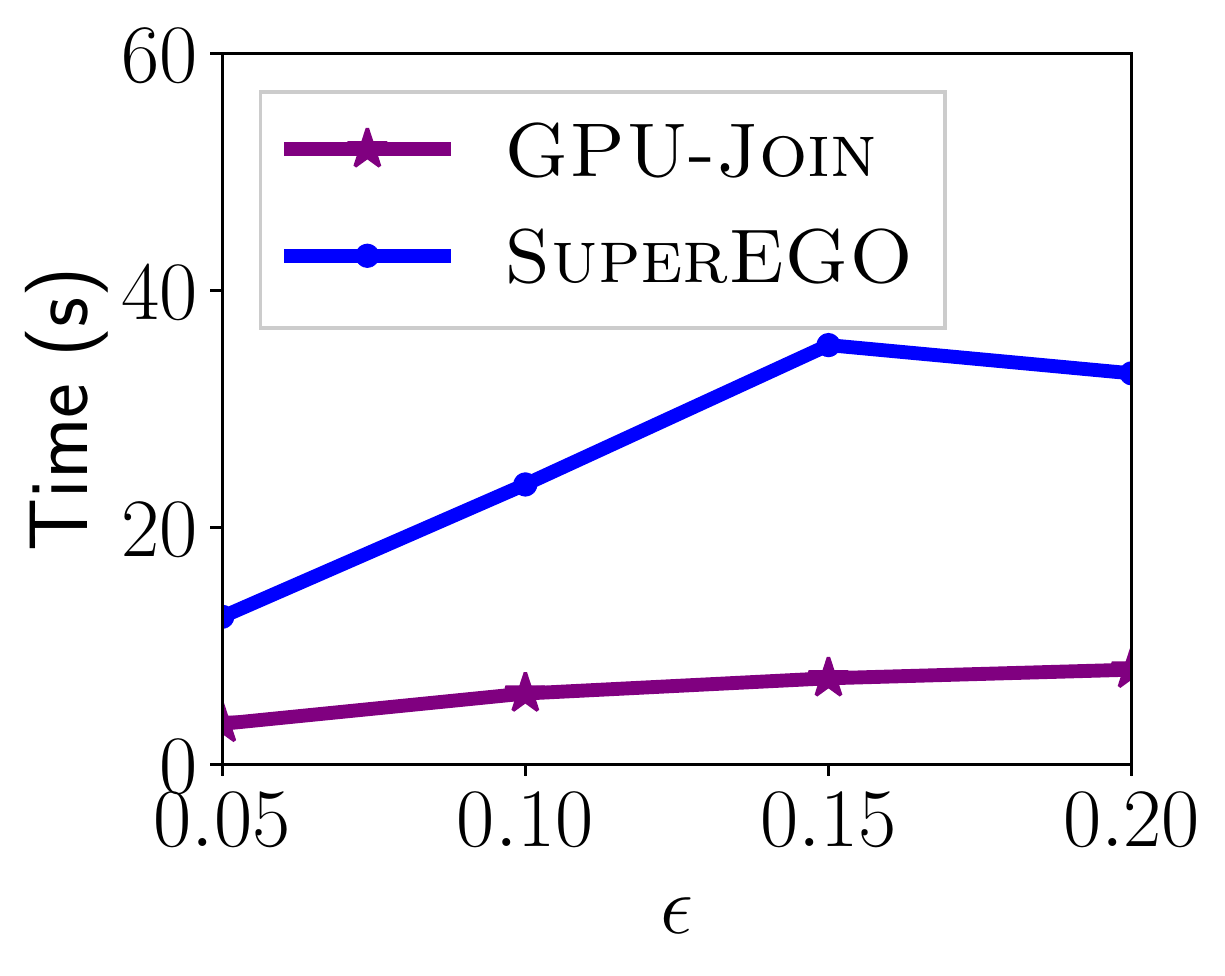}
    }            
    \caption{Response time vs. $\epsilon$ on the real-world datasets used in \ego~\cite{kalashnikov2013}. $k=6$ dimensions are indexed across all datasets. Rounded values of $S_D$ in the plots are in the range (a) 4--26k, (b) 3--1.1k, and (c) 14k--62k.}
   \label{fig:real_world_ego}
\end{figure}




Figure~\ref{fig:real_world_ego} plots response time vs. $\epsilon$ on \colorhist, \layout, and \cooc. We find that across all datasets \gpu tends to outperform \ego (recall that \ego uses 16 cores/threads) on all experimental scenarios. In Figure~\ref{fig:real_world_ego}~(b) we see that \gpu and \ego have nearly identical performance when $\epsilon\leq0.15$, but when $\epsilon>0.15$, the performance of the approaches diverge. Also, note that in Figure~\ref{fig:real_world_ego}~(c) the performance of \ego increases and then decreases. This is because by $\epsilon=0.15$, nearly all neighbors are within $\epsilon$ of each other on this dataset. Since \ego uses dimensionality reordering as a function of $\epsilon$, it is possible that performance decreases with small increases in $\epsilon$ due to the filtering effects of the algorithm. 

The reason the performance of \gpu does not degrade significantly with $\epsilon$ is because these datasets 
are relatively small. 
There is little work for the GPU to execute, and the GPU's resources are not fully saturated. Thus, as $\epsilon$ increases, the response time does not increase in the same manner as \ego. Furthermore, these datasets could nearly (or entirely) fit in the L3 cache of a modern CPU. Thus, to fully leverage the GPU and observe the effects of our optimizations, we need to execute \gpu on larger workloads. However, our comparison of \gpu and \ego on these datasets used by Kalashnikov~\cite{kalashnikov2013} demonstrates that \gpu outperforms the state-of-the-art algorithm.   

For the reader that would like to compare our work to \ego, we note that the selectivity values (the average number of neighbors), $S_D$, in the corresponding plots provided by Kalashnikov~\cite{kalashnikov2013} are consistent for \colorhist and \layout, but not for \cooc. We cannot determine why there is an inconsistency, but we suspect that it is due to differences in dataset normalization.

\subsection{Index Dimensionality Reduction and Reordering}\label{sec:eval_dim_reduction_reorder}
Recall from Section~\ref{sec:indexkdim} that, by indexing $k<n$ dimensions, we have a less discriminating search for points, which yields larger candidate sizes that need to be filtered using Euclidean distance calculations. 
However, this is needed to reduce the overhead of searching for high-dimensional data (recall that the total number of grid cells searched for a query point is $3^n$). We also reorder the points to exploit variance in the dimensions of the data (Section~\ref{sec:reorder}). This has the effect of indexing on the dimensions that most improve the filtering power of the index.


\begin{figure}[tp]
\centering
  \subfloat[\susy (18-D), $\epsilon=0.01$]{
        \includegraphics[width=0.22\textwidth, trim={0.5cm 0 0.5cm 0}]{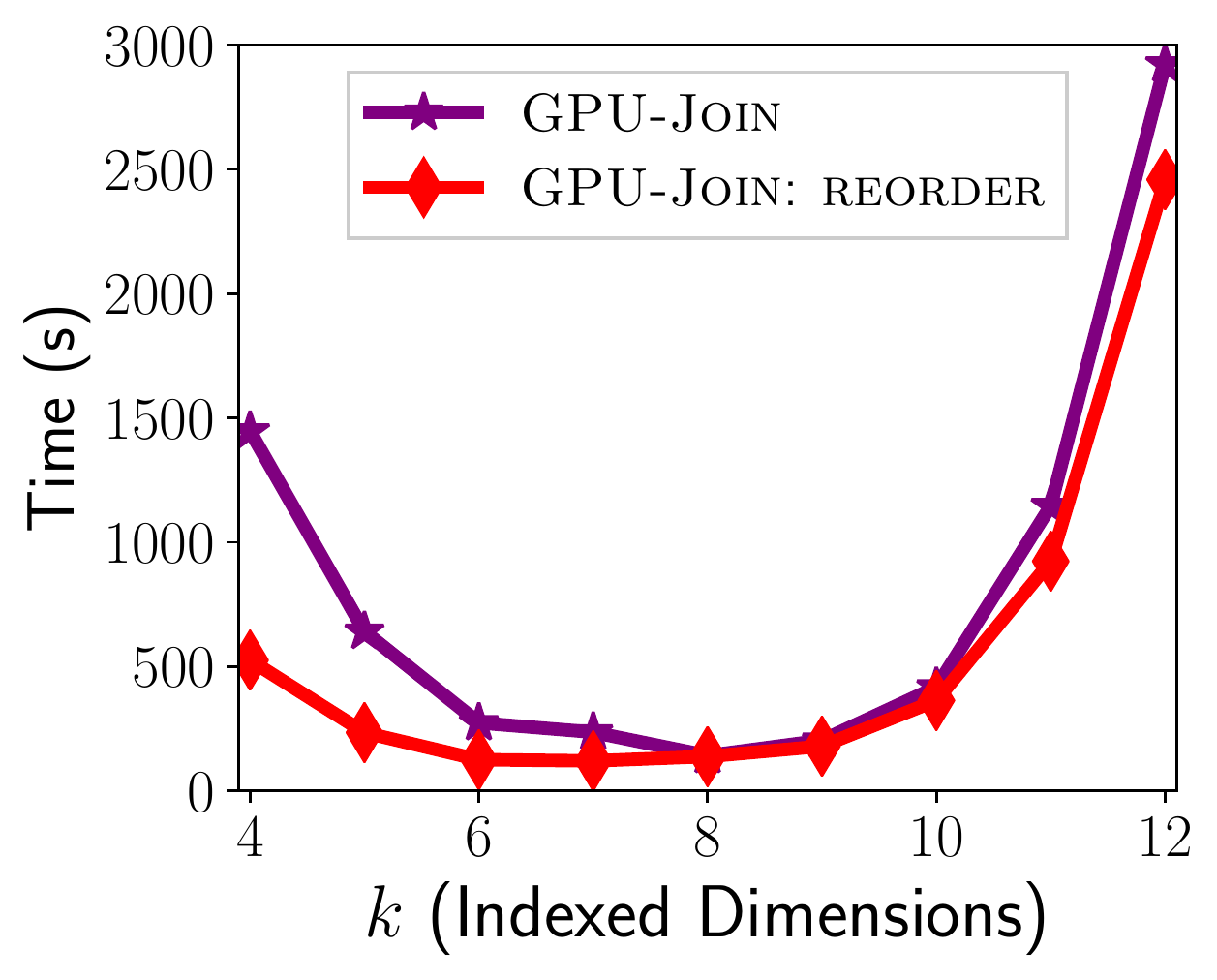}
        }
      \subfloat[\songs (90-D), $\epsilon=0.005$]{
	      \includegraphics[width=0.22\textwidth, trim={0.5cm 0 0.2cm 0}]{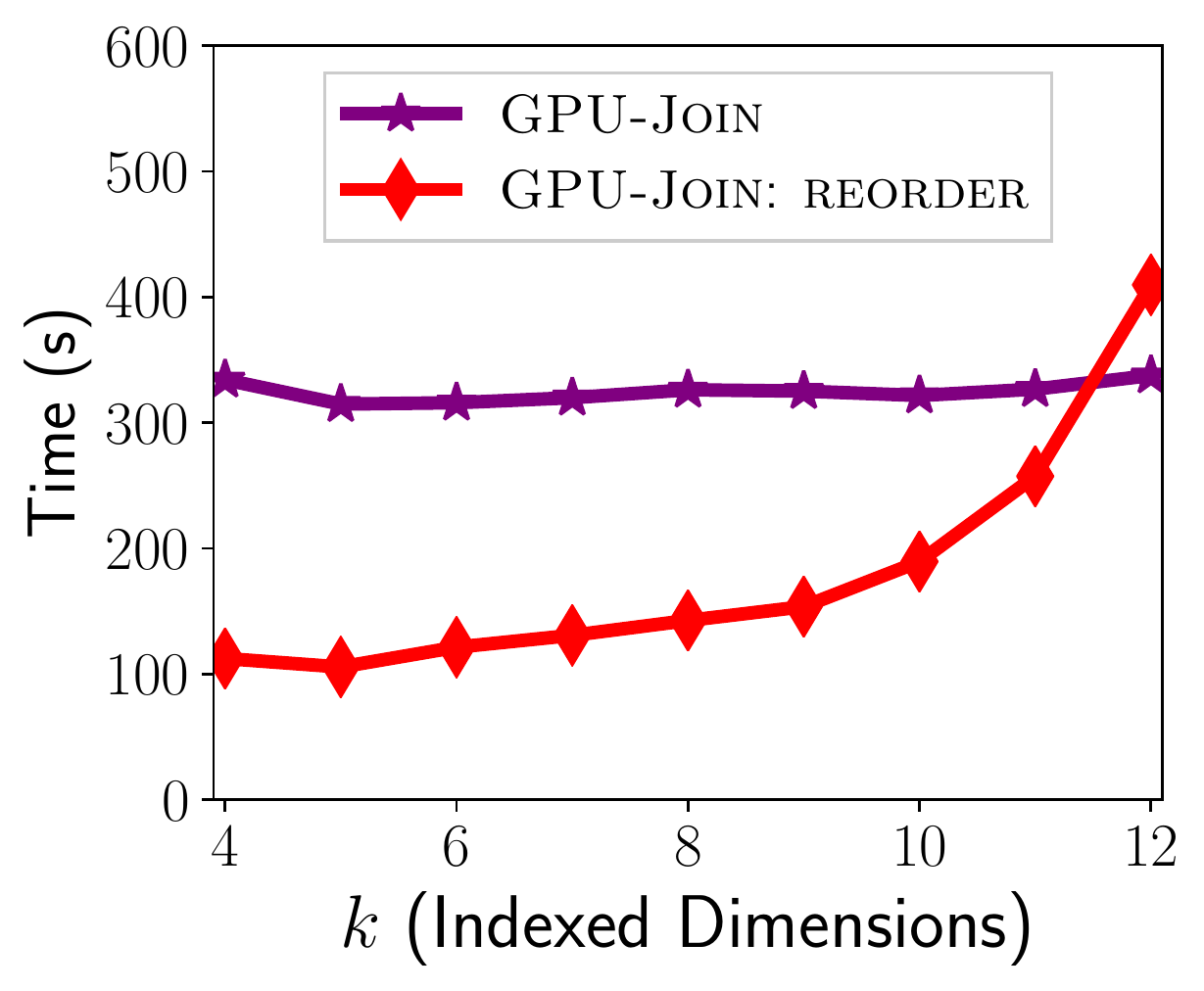}
        }    
    \caption{Response time vs. indexed dimension, $k$.}
   \label{fig:k_selection}
\end{figure}


Figure~\ref{fig:k_selection}~(a) shows the response time vs. $k$ on \susy with and without using \reorder. In both cases, we see that performance degrades when too few or too many dimensions are indexed due to increased point comparisons or search overhead, respectively. However, when we reorder the data by variance, the response time is significantly reduced, particularly for small $k$. Since the data is indexed in the first $k$ dimensions when we do not use \reorder, it is possible that a given dataset will still exhibit good performance if the first $k$ dimensions, \emph{by chance}, have a large amount of variance. However, if they do not, performance will significantly degrade. The worst possible scenario would be when the variance is so small in the first $k$ indexed dimensions that all of the points fall within a single grid cell, resulting in $O(n^2)$ distance comparisons.  In such a case, using \reorder can significantly improve the ability of the index to prune the search for points within $\epsilon$.  Figure~\ref{fig:k_selection}~(b) shows the same plot for the \songs dataset. This is an example where the first $k\lesssim12$ dimensions have low variance, thereby generating a grid with few cells and low index filtering power.  We note that when $k>10$ the response time when using \reorder exceeds the time when not using the approach. This is because while \reorder exploits variance to improve the index filtering power, it also increases the number of cells in the index, increasing the number of cells to consider when performing index searches. 
However, for either \susy or \songs, \reorder significantly reduces the response time when $3\leq k\leq 8$, which is a large range from which to select $k$. Thus, hereafter we simply select $k=6$ dimensions, but describe how to select $k$ in the next subsection.

\subsection{Selecting the Number of Dimensions to Index}
When indexing on fewer dimensions (i.e., $k < n$), there is an increase in the number of distance calculations, but fewer cells are searched. Thus, we describe a method to select a good value of $k$. Real-world and skewed datasets in high dimensions do not allow us to use aggregate metrics to select $k$, e.g., the average number of cells searched, the average number of point comparisons, and the average number of neighbors per point. Thus, we cannot utilize analytical methods to estimate the amount of work needed to perform the self-join. Instead, we rely on a sampling technique that we execute when estimating the total result set size needed for batching (Section~\ref{sec:batching}). Sampling the datasets yields an estimate of selected metrics, and can be computed in a small fraction of the total response time of \gpu. When computing the index, we store the number of non-empty grid cells, denoted as $|G|$. For a given value of $k$, we execute \gpu for a fraction $f$ of the data points, and record the number of point comparisons (points that are tested to be within $\epsilon$ of each other), denoted as $\mu$. The estimated total number of memory operations to search all of the cells is as follows: $|D|3^k \rm{log_2}(|G|)$, where for each point in the dataset, we search the adjacent cells, and perform a binary search to find the non-empty cells that exist in the index. The estimated total number of memory operations needed for the distance comparisons is simply $\mu(1/f)$. We can select a good value of $k$ by comparing the total number of memory operations for: $(i)$ searching whether the cells exist; and, $(ii)$ distance comparisons.

\begin{figure}[htp]
\centering
        \includegraphics[width=0.3\textwidth, trim={0.4cm 0 0.6cm 0}]{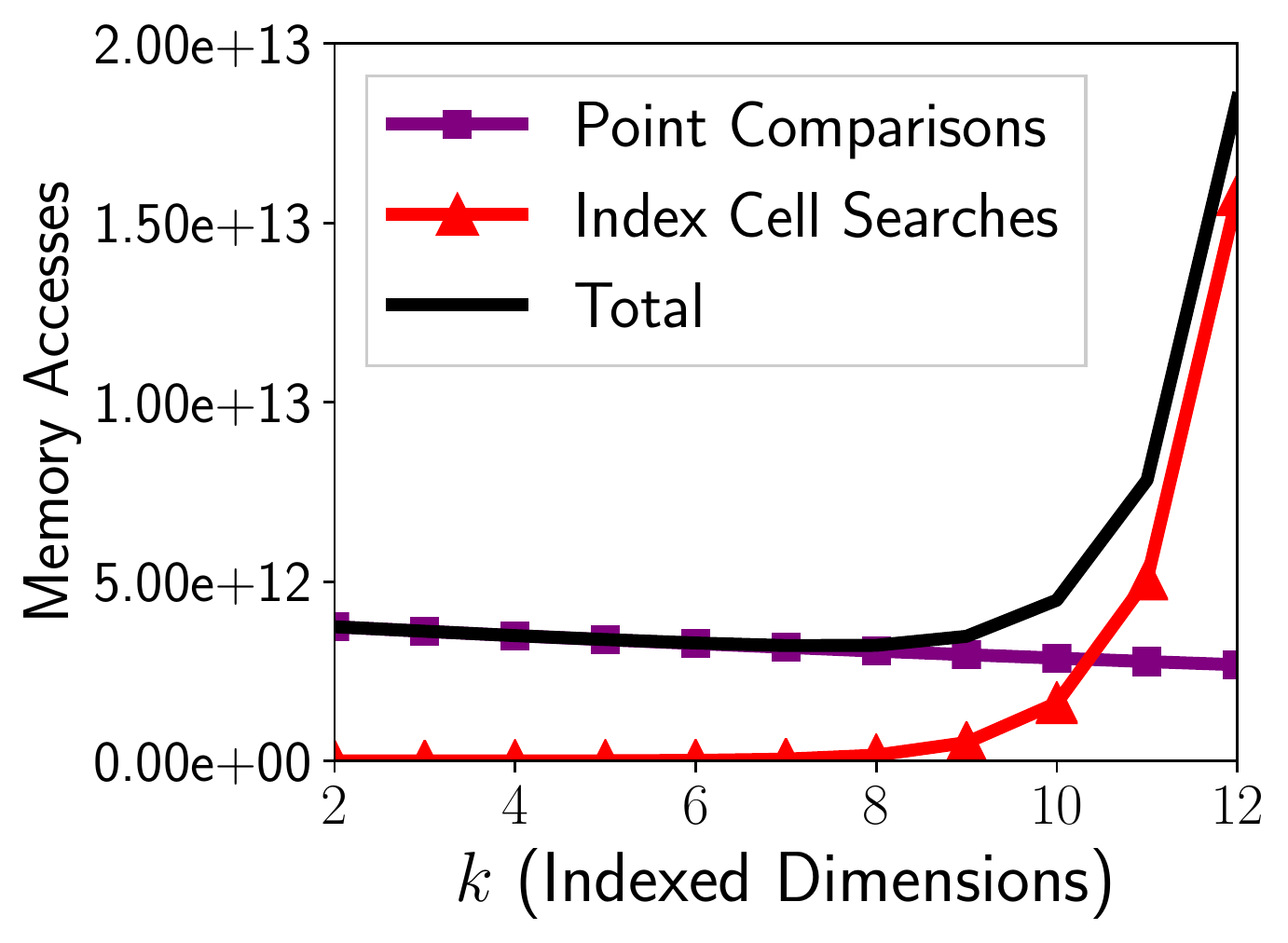}
    \caption{The number of index search and point comparison memory accesses.}
   \label{fig:memory_ops}
\end{figure}


Figure~\ref{fig:memory_ops} plots the number of memory operations vs. $k$ on the \dsynthah dataset. As expected,  we observe a reduction in the number of distance comparison memory operations with increasing $k$, indicating that indexing on more dimensions reduces the number of distance comparisons. While the number of memory operations associated with the cell searches is relatively low for $k\leq8$, the exponential increase in the number of adjacent cells with increasing $k$ makes indexing $k>10$ significantly degrade performance. We note that since we use an exponentially distributed dataset, there are many point comparisons, as the points are strongly co-located. For datasets that exhibit a more uniform distribution, we expect that varying $k$ will more have a larger impact on the number of point comparisons.
In this case, a different value of $k$ may be more appropriate than the value of $k$ for \dsynthah. 

Regardless of the dataset used, it is clear that indexing $k>10$ is likely to degrade performance; however, we observe that in terms of the number of memory accesses needed for index cell searches, in practice we can select a value of $k$ within a fairly large range, roughly within $k\leq10$ without a significant performance penalty. This demonstrates that we do not need excessive parameter tuning of $k$ to achieve good performance.

\subsection{Performance on Larger Datasets}
We execute \gpu on larger real-world and synthetic datasets (the latter being the worst case for \gpu).

\subsubsection{Real World Datasets}
\begin{figure}[tp]
\centering
\subfloat[\susy (18-D)]{
        \includegraphics[width=0.22\textwidth, trim={0.6cm 0 0.6cm 0}]{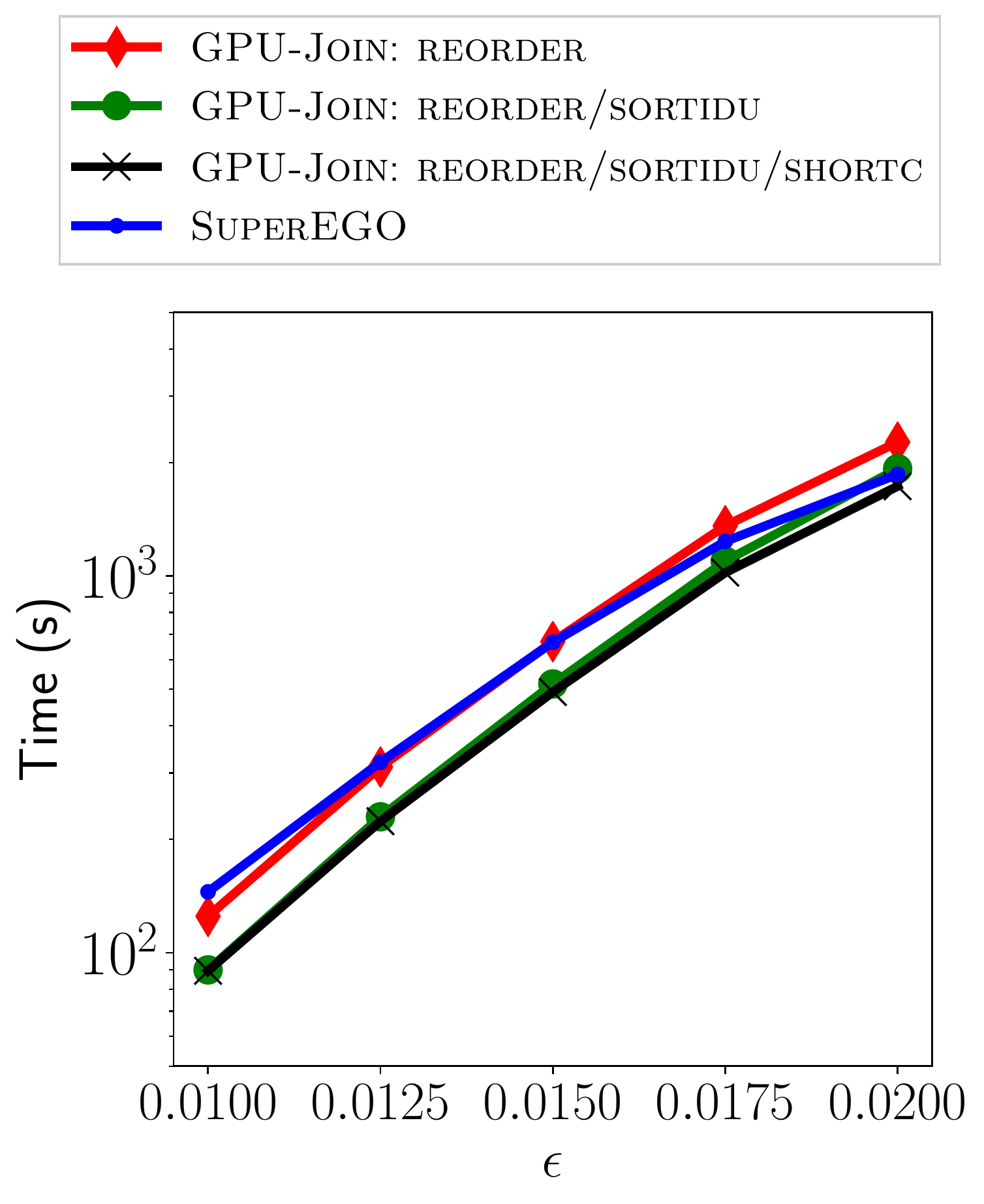}
    }          
\subfloat[\songs (90-D)]{
        \includegraphics[width=0.22\textwidth, trim={0.4cm 0 0.6cm 0}]{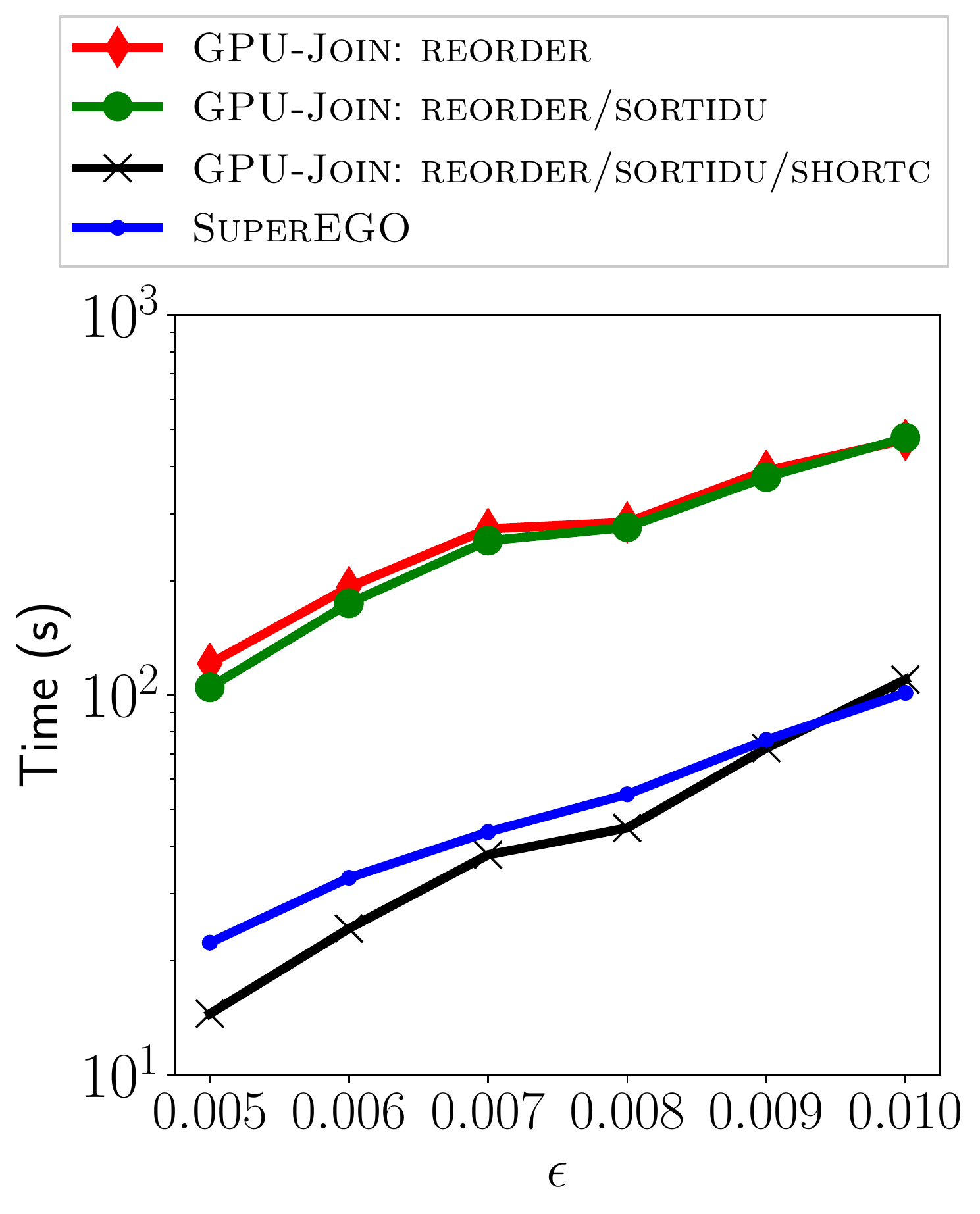}
    }          

    \caption{Response time vs. $\epsilon$. $k=6$ dimensions are indexed. Rounded values of $S_D$ in the plots are in the range (a) 5--781, and (b) 4--1.9k.}
   \label{fig:real_world_larger}
\end{figure}

Figure~\ref{fig:real_world_larger}~(a) plots the response time vs. $\epsilon$ on the \susy dataset.  Note that the \susy dataset is two orders of magnitude larger ($|D|=5\times10^6$) than those used in Figure~\ref{fig:real_world_ego}.  
Results indicate that, for the \susy dataset, \sortidu reduces response time by a reasonable margin (e.g., at $\epsilon=0.01$, using \sortidu and \reorder is 38\% faster than \reorder alone), though the \scircuit optimization has a negligible effect. Using all optimizations, \gpu outperforms \ego across all values of $\epsilon$, yielding speedups  between 1.07$\times$ ($\epsilon=0.02$) and 1.61$\times$ ($\epsilon=0.01$).



Figure~\ref{fig:real_world_larger}~(b) plots the response time vs. $\epsilon$ on the \songs dataset, which is the highest dimensional dataset that we use. We see that using \sortidu reduces the response time at lower values of $\epsilon$. In contrast to the \susy dataset (Figure~\ref{fig:real_world_larger}~(a)), \scircuit yields a significant reduction in response time on the \songs dataset due to its much higher dimensionality. We note that \ego also employs an optimization that short circuits the distance calculation.
We find that \gpu outperforms \ego across all values of $\epsilon$ with the exception of $\epsilon=0.01$. Thus, the speedup (or slowdown) over \ego ranges from 1.53$\times$ ($\epsilon=0.005$) to 0.92$\times$ ($\epsilon=0.01$). The slight slowdown on \songs at $\epsilon=0.01$ indicates that \ego is competitive with \gpu under some experimental scenarios.             



\subsubsection{Synthetic Datasets}
We use synthetic datasets to understand the performance of \gpu when the approach cannot exploit the dimensionality reordering optimization (Section~\ref{sec:reorder}). Under these conditions, it is not possible to reorder these data to improve the filtering power of the index because the variance is nearly the same in each dimension. This data distribution may be common in real-world scenarios if all of the features are drawn from the same statistical distribution. We utilize synthetic datasets with an exponential distribution. If we were to use a uniform, or even a normal distribution with a moderate variance, we will be unable to find many (or any) neighbors within $\epsilon$ because the points will be too far away from each other in high dimensions. Thus, using an exponential distribution ensures that we will find on average a reasonable number of neighbors for a given $\epsilon$, similarly to finding neighbors in the high dimensional real-world datasets.

\begin{figure}[tp]
\centering

\subfloat[\dsynthah (16-D)]{
        \includegraphics[width=0.22\textwidth, trim={0.4cm 0 0.4cm 0}]{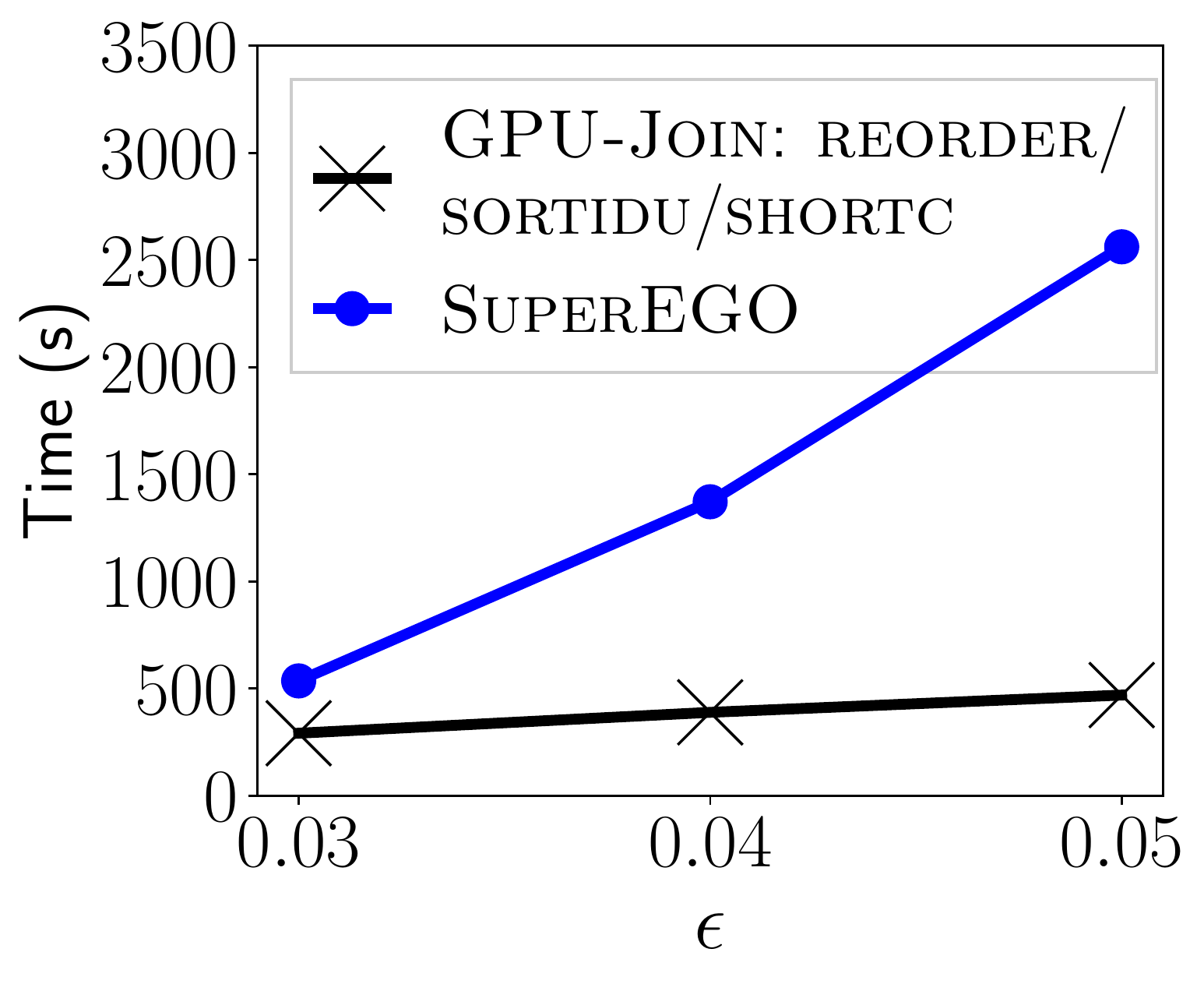}
    }      
\subfloat[\dsynthai (32-D)]{
        \includegraphics[width=0.22\textwidth, trim={0.6cm 0 0.6cm 0}]{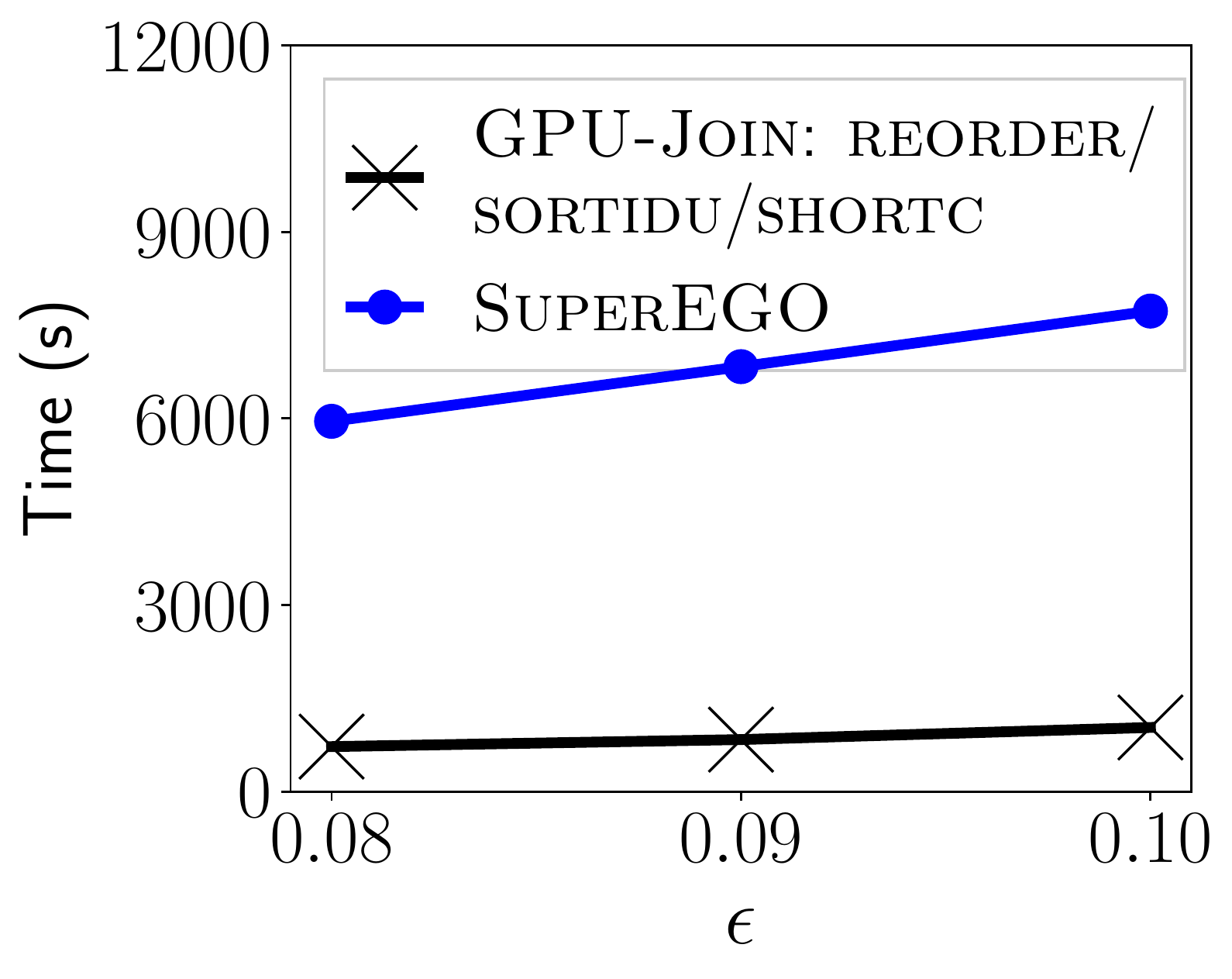}
    }     

\subfloat[\dsynthaj (64-D)]{
        \includegraphics[width=0.22\textwidth, trim={0.4cm 0 0.4cm 0}]{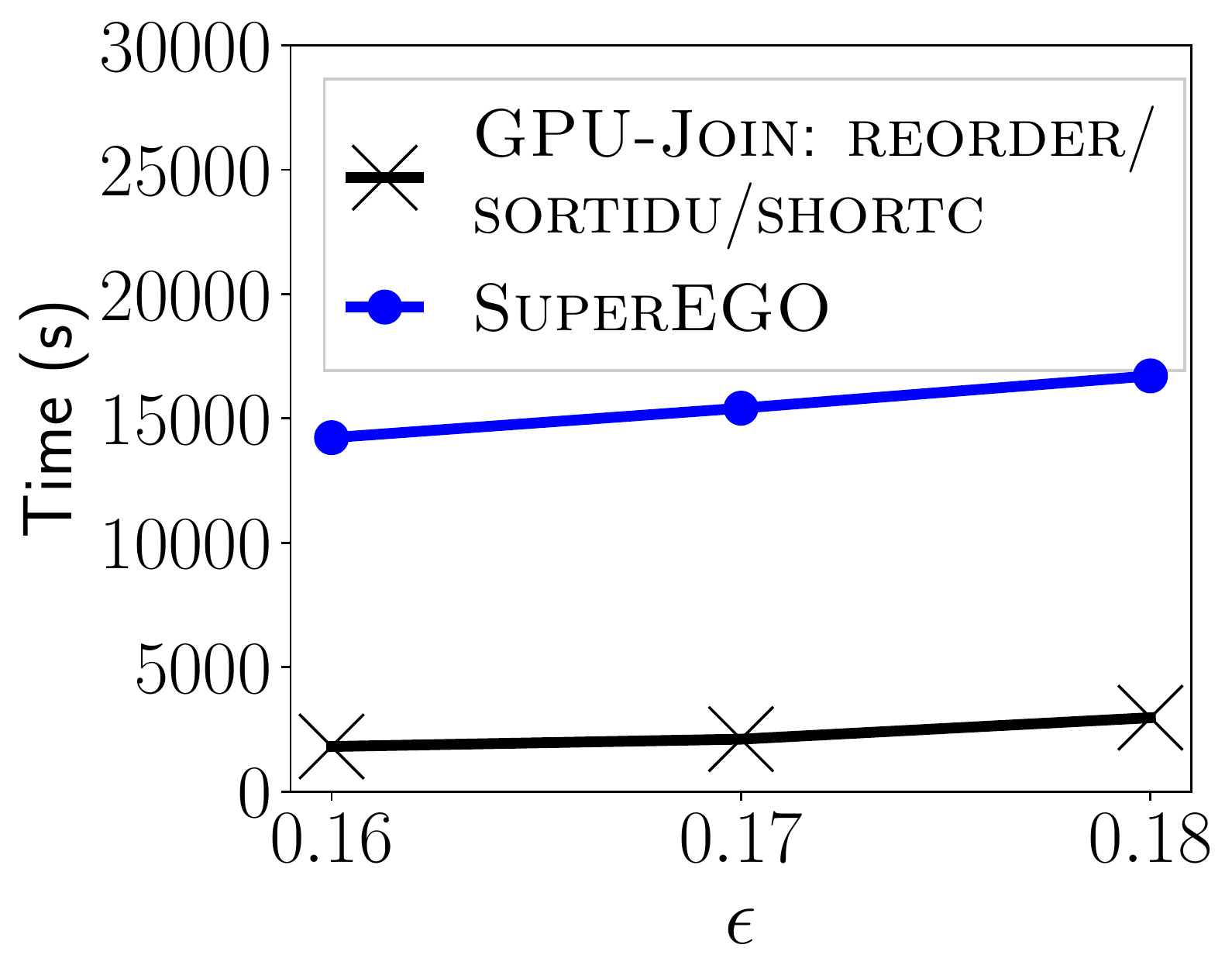}
    }     
    \caption{Response time vs. $\epsilon$ on the synthetic datasets. 
$k=6$ dimensions are indexed and \gpu is configured using all optimizations. Rounded values of $S_D$ in the plots are in the range (a) 4--1.2k, (b) 31--1.4k, and (c) 132--2.3k.}
   \label{fig:synthetic}
\end{figure}

Figure~\ref{fig:synthetic} plots the response time vs. $\epsilon$ on 16, 32, and 64-D synthetic datasets. We observe that \gpu outperforms \ego on all scenarios, with the smallest performance gain on the smallest workload (Figure~\ref{fig:synthetic}~(a), $\epsilon=0.03$), where the speedup is 1.84$\times$. The largest speedup is 8.25$\times$ on \dsynthai with $\epsilon=0.08$. Both \ego and \gpu exploit the statistical properties of the data. However, these results indicate that the index search performance of \ego is more sensitive to the dimensionality reordering technique~\cite{kalashnikov2013} than the \reorder technique used by \gpu. In cases where the variance is similar across dimensions, \gpu is likely to significantly outperform \ego.

\section{Scaling to Multi-GPU Systems}\label{sec:scalability_pathways}
Further improving single-GPU performance through additional algorithmic innovations are unlikely to yield large performance gains (there are certainly improvements to \gpu, but likely none that can achieve order-of-magnitude reductions in the time-to-solution). The self-join, particularly in high dimensions, is an expensive operation.  For instance, on the \susy dataset in Figure~\ref{fig:real_world_larger}~(a) at $\epsilon=0.02$, the response time of \gpu with all optimizations is 1729~s. In this section, we discuss potential strategies to efficiently apply \gpu to  multi-GPU and distributed-memory systems.

In multi-GPU systems, one issue is contention on the PCIe bus, assuming a configuration where all GPUs communicate over PCIe (new interconnects such as NVLink will reduce contention~\cite{2017Foley}).  For distributed-memory systems, a key bottleneck is communication between compute nodes~\cite{Lynch:distributed}. Thus, the communication requirements in either system may degrade performance. Another bottleneck in either configuration is load balancing; ideally all nodes/GPUs have similar workloads and finish at approximately the same time.
If there is a large load imbalance, the performance will not scale well as we increase the number of nodes/GPUs.

\subsection{Single-Node Multi-GPU Overheads}
\label{sec:scalability-single-node}
As mentioned above, load balancing and communication overheads are a concern for multi-GPU systems. With regards to high-dimensional \gpu performance, we demonstrate that the overhead of using a single-node, multi-GPU system is negligible.  
We quantify the single-GPU host-side overheads as follows. We execute \gpu on our single-GPU platform with a single batch/stream (i.e., not overlapping data transfers and host-side operations), and record the total compute-only time needed to execute the GPU kernel across all of the batches. Table~\ref{tab:hostOH} shows the compute-only and total response times, and the percentage of time spent performing host-side operations (data transfers and overheads) on \dsynthah and \susy. We find that on \susy and \dsynthah, only 0.69\% and 1.8\%, respectively, of the total response time is needed for host-side and data transfer operations. Note that, since this result is obtained using a single stream, the actual communication overhead is much less due to pipelining (see Figure~\ref{fig:pipeline_batches}).  
Therefore, overheads associated with PCIe and Host-GPU communication are negligible compared to overall \gpu response time.  This indicates that performance will scale well on single-node, multi-GPU systems or systems that require little inter-node communication (i.e., if $|D|$ does not exceed GPU global memory).  Thus, if we can achieve a balanced workload across all GPUs, the performance of \gpu will scale well on multi-GPU systems.



\begin{table}[htp]
\centering
\caption{Compute-only time compared to the time performing other host-side operations (overhead).}\label{tab:hostOH}
\begin{tabular}{|c|c|c|c|c|} \hline
Dataset  &$\epsilon$&Compute (s) &Total (s)& Overhead\\ \hline
\dsynthah &0.05 &460.58&469.00&1.8\%\\\hline
\susy     &0.02    &1717.26&1729.24&0.69\%\\\hline
\end{tabular}
\end{table}

\subsection{Balancing Workloads: Entity Partitioning}\label{sec:scalability_pathway_entity}
We propose an entity partitioning approach to achieve
a balanced workload among nodes.
We denote each processing element (node) as a single GPU, $p_k$, where $k=0,\dots,|p|-1$, and $|p|$ is the total number of GPUs. 
We considered a spatial partitioning strategy, but found that it does not ensure a balanced workload across GPUs.

In our experimental evaluation of \gpu on our single-GPU platform, the size of the dataset, $|D|$, does not exceed main memory on the host or the GPU's global memory capacity. Since the datasets never exceed global memory in our experimental evaluation, and $|D|\ll|R|$, we store the entire dataset on each $p_k$ once. Recall that each $p_k$ is assigned a uniform selection of the total dataset to query. Formally, we denote the total number of entity partitioned query batches as $N_b$. We define a set of entity partitioned query point sets as $Q_l$, where $l=0,\dots,N_b-1$, where the size of each query set is simply $|Q_l|=|D|/N_b$ (assuming $|D|$ evenly divides $N_b$). Let us assign the $N_b$ query sets to $p_k$ in a round robin fashion, where GPU $p_k$ is assigned $Q_l$ to perform the join operation, $Q_l \bowtie_\epsilon D$, if $l~mod~|p| = k$.  To ensure good load balancing, we select $N_b$ such that $N_b~mod~|p|=0$, and therefore, all GPUs have the same number of query sets to process.      

\begin{figure}[t]
\centering
\subfloat[\dsynthah]{
        \includegraphics[width=0.22\textwidth, trim={0.5cm 0 0.5cm 0}]{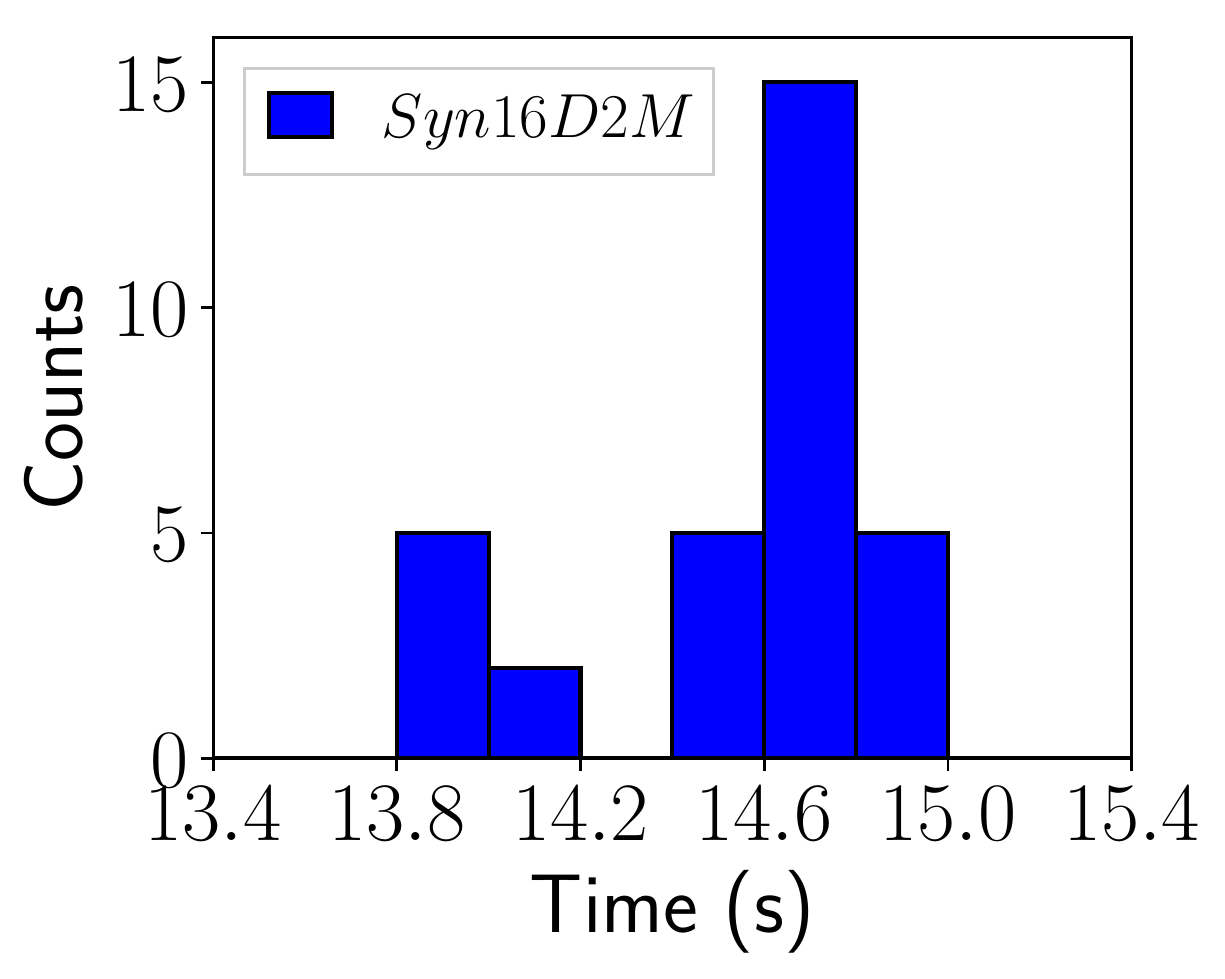}
    }      
\subfloat[\susy]{
        \includegraphics[width=0.22\textwidth, trim={0.5cm 0 0.5cm 0}]{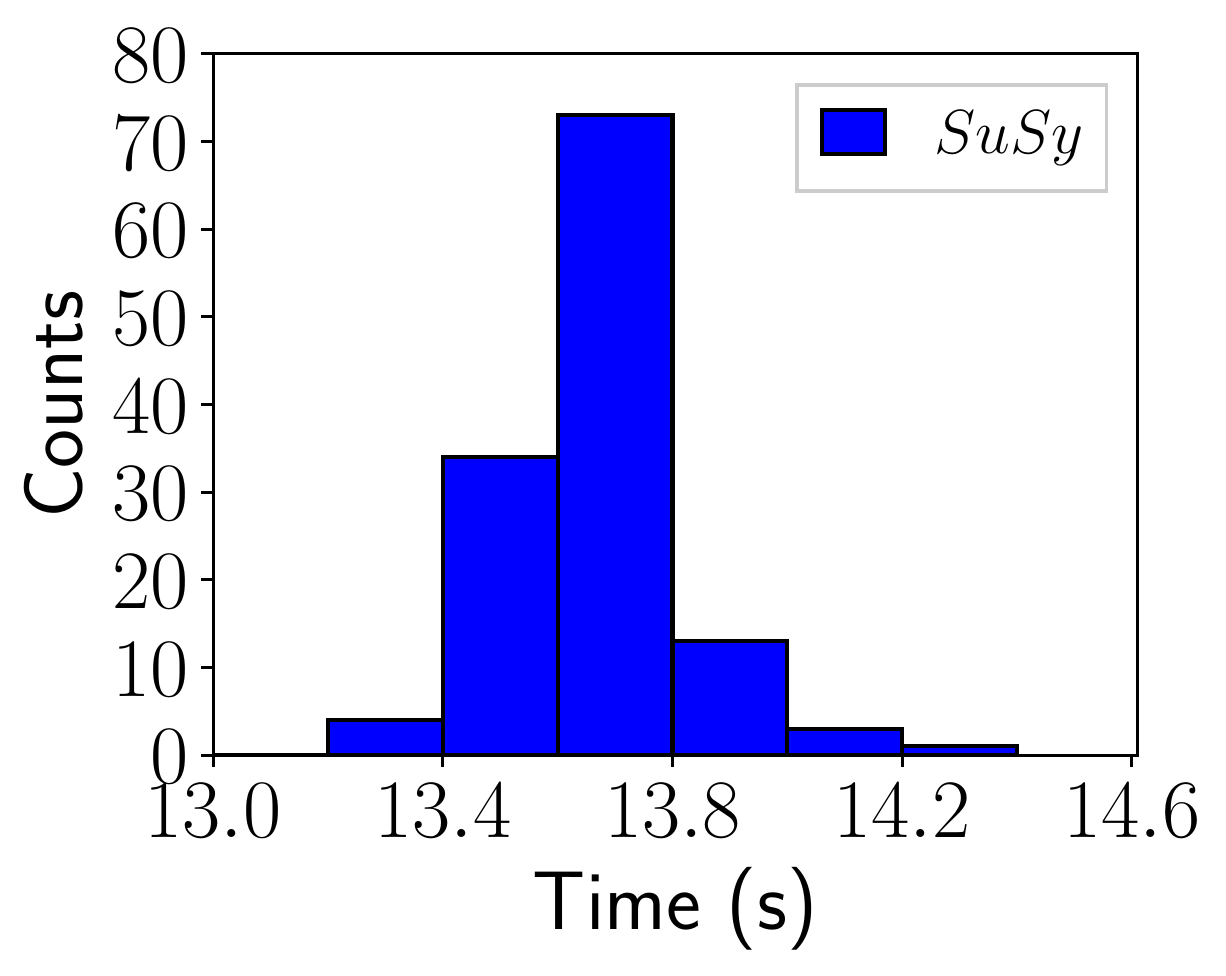}
    }   

    \caption{Histogram of the computation time of entity partitioned batches in 0.2~s bins for (a)~\dsynthah with $N_b=32$, and $\epsilon=0.05$; (b)~\susy with $N_b=128$, and $\epsilon=0.02$.}
   \label{fig:histogram}
\end{figure}

The scalability of \gpu on multi-GPU systems is largely a function of load balancing. Thus, to ascertain the expected scalability of \gpu on a distributed-memory system, we consider the computation time of the work performed by each node.  Figure~\ref{fig:histogram}~(a)--(b) plots a histogram of the computation times (excluding all other host-side bottlenecks and overheads as discussed in Section~\ref{sec:scalability-single-node}) of performing the join between each query set and the dataset as executed separately on our single-GPU platform.  
We generate $N_b=32$ and $N_b=128$ query sets for \dsynthah and \susy, respectively (the datasets are different sizes, so we generated a different number of query sets for each).   We find that on the \dsynthah and \susy datasets, the minimum (maximum) kernel execution times are: 13.88~s (14.95~s) and 13.36~s (14.34~s), respectively, indicating a small variance among the computation time of each query batch. This small variance demonstrates the efficacy of the entity partitioning strategy to achieve a balanced workload. Furthermore, if these query sets were to be assigned to multiple GPUs, we would expect that each GPU would finish its computation at roughly the same time (assuming $N_b~mod~|p|=0$).

\begin{figure}[t]
\centering
\subfloat[\dsynthah]{
        \includegraphics[width=0.22\textwidth, trim={0.5cm 0 0.5cm 0}]{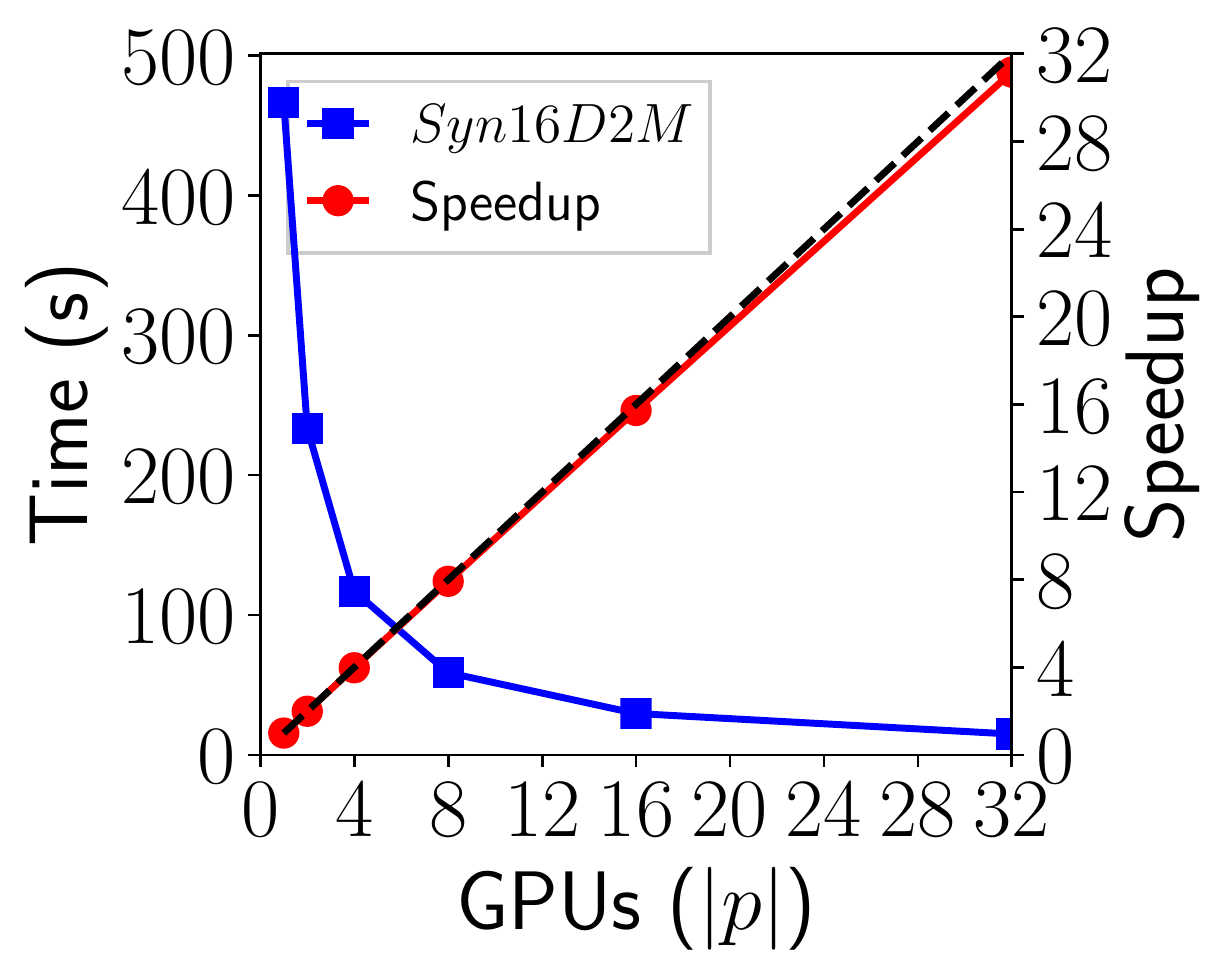}
    }      
\subfloat[\susy]{
        \includegraphics[width=0.22\textwidth, trim={0.5cm 0 0.5cm 0}]{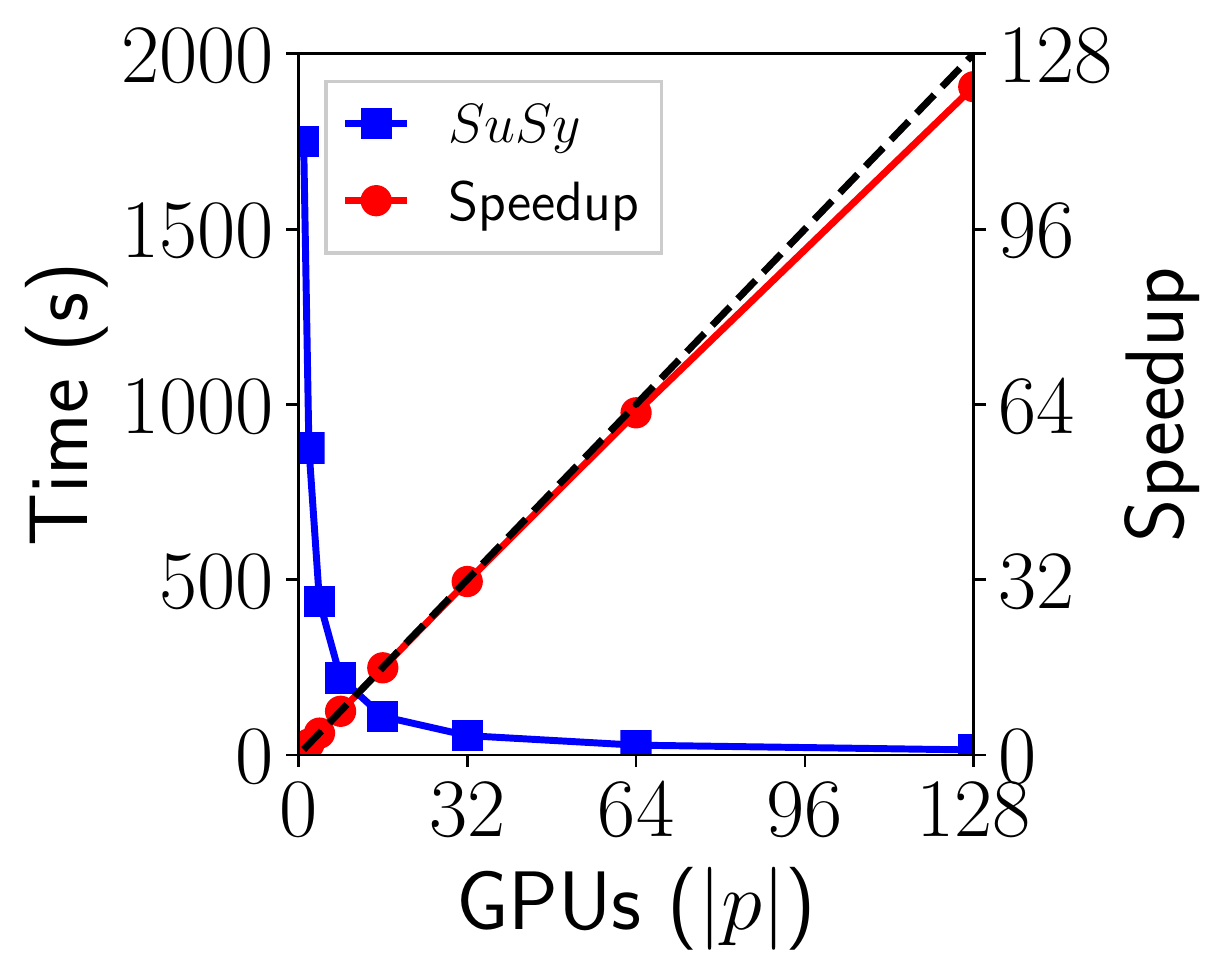}
    }          

    \caption{Scalability on (a)~\dsynthah, and (b)~\susy, using the response time measurements summarized in Figure~\ref{fig:histogram}.}
   \label{fig:scalability}
\end{figure}

To understand the potential scalability of \gpu, we simulate the expected response time if we executed the query batches in Figure~\ref{fig:histogram} on $|p|$ GPUs. We ignore host-side overheads, as we demonstrated that they are negligible in Section~\ref{sec:scalability-single-node}. As described above, we generate a fixed number of query batches ($N_b$) and assign them to the $|p|$ GPUs in a round-robin fashion. Figure~\ref{fig:scalability} plots the response time and speedup of \gpu for varying $|p|$ (and fixed $N_b$). 
We find that the entity partitioning strategy achieves near-ideal scalability. Assuming that $|D|$ does not exceed the global memory capacity of a single GPU, this can be executed on a distributed-memory system without any communication overhead.  Each node independently computes its assigned query batches ($Q_l$) from its corresponding MPI rank and communicator size (assuming an MPI implementation).
Thus, we expect that distributed-memory self-joins will exhibit good scalability when $D$ can be replicated on each GPU.
However, on multi-node systems where $|D|$ exceeds global memory, inter-node communication may degrade performance and reduce scalability.

\subsection{Extending to Very Large Datasets}
In this previous section, we demonstrated that \gpu achieves excellent scalability when the dataset is duplicated on each GPU. However, if the dataset exceeds the global memory capacity of a single GPU, we must partition and distribute the data among the different nodes in the system.  Thus, we outline a partitioning and communication strategy that allows the dataset to be distributed across multiple compute nodes such that we can perform \gpu on larger datasets.  

We assume the input set, $D$, is initially evenly distributed among all $|p|$ nodes (i.e., each node has a uniform entity partitioned selection of $|D|/|p|$ data points).  We call this initial set of points the \emph{query points}, and we refer to the set of query points of node $p_k$ as $Q_k$, where $|Q_k|=|D|/|p|$.  Each node makes a copy of this set and
calls it the entry data points, $E_k$ (entry meaning a set of entries in the index).  To solve the global self-join, each node computes the join from
its set of query points to every other data point in $D$.  To do this, each node $p_k$ starts by performing an independent join operation between $Q_k$ and $E_k$, which
we denote $Q_k \bowtie_\epsilon E_k$.   Each node then sends its set $E_k$ to the next node, $p_{(k+1)mod|p|}$ and receives a new entry set from processor $p_{(k-1)mod|p|}$.  Node $p_k$ then performs $Q_k \bowtie_\epsilon E_{(k-1)mod|p|}$ and, again, passes the entry data set to node $p_{(k+1)mod|p|}$.  
This continues for $|p|-1$ rounds, at which point all nodes have compared their query points to each subset of the entity partitioned data. The result of all the independent join operations constitutes the total self-join. 

While our partitioning and communication strategy allows us to execute \gpu on larger datasets with a balanced workload across a distributed-memory system, it comes at the cost of additional communication overheads.  
Specifically, each communication phase requires transmitting a total of $|p|\times|D|/|p|=|D|$ elements between nodes, and in total there are $|p|-1$ communication phases. Thus, the total number of data elements communicated is $(|p|^2-|p|)\times|D|/|p|=(|p|-1)|D|$, with each node sending and receiving a total of $|D|-(|D|/|p|)$ data elements.

\begin{figure}[t]
\centering
\subfloat[\dsynthah]{
        \includegraphics[width=0.22\textwidth, trim={0.5cm 0 0.5cm 0}]{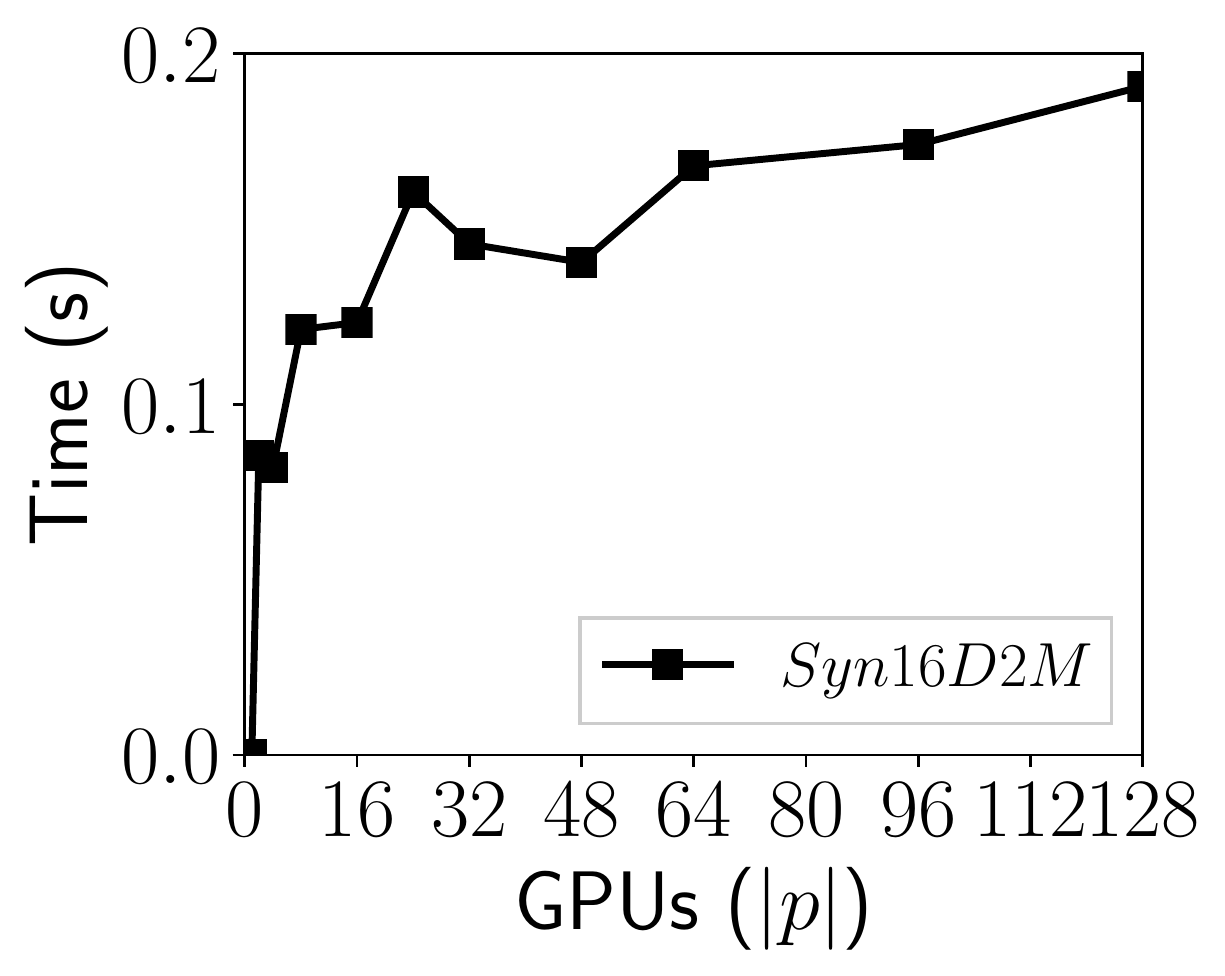}
    }      
\subfloat[\susy]{
        \includegraphics[width=0.22\textwidth, trim={0.5cm 0 0.5cm 0}]{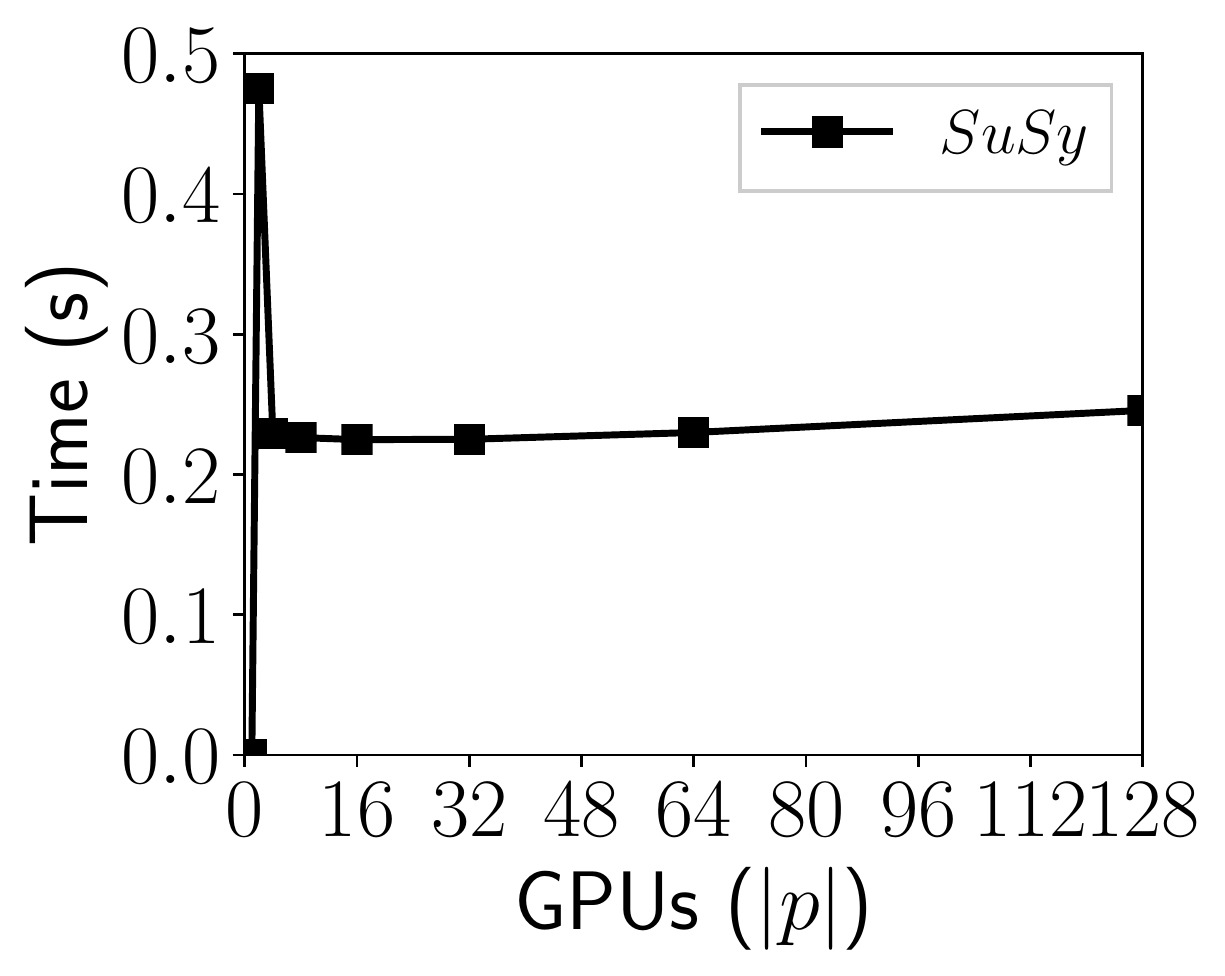}
    }                  
    \caption{Communication cost of the entity partitioning strategy on up to $|p|=128$ nodes for (a)~\dsynthah; and (b)~\susy.}
   \label{fig:distributed_communication}
\end{figure}

To determine the cost of communication overhead, we assume a scenario where there is one GPU per node on a distributed-memory cluster.  
We measure communication overhead by implementing the inter-node communication required by the data partitioning and communication strategy outlined above using MPI 4.1 as executed on up to 128 nodes in a cluster with a 40Gbit/s InfiniBand interconnect.
We compute the time needed to partition the data and perform inter-node communication, assuming $D$ is initially distributed across $|p|$ nodes. We assume a Bulk Synchronous Parallel (BSP)~\cite{valiant1990bridging} communication pattern, where at each global superstep the data is transferred between nodes and then all nodes synchronize before the next round of GPU computation can begin (although we do not incorporate computation). Thus, the time of each communication phase is measured from the start of the communication phase until last node completes communicating. 
Let $t_a$ be the time needed to communicate at phase $a$, where $a=1,\dots,|p|-1$ (since there are $|p|-1$ communication phases). Thus, the total communication time, $t$, is 
\begin{eqnarray}
t = \sum_{i=1}^{|p|-1}t_i.
\end{eqnarray}
Figure~\ref{fig:distributed_communication}~(a)--(b) plots the total communication time of the distributed entity partitioning strategy on the \dsynthah and \susy datasets, respectively, on up to 128 nodes on the cluster. Communication cost is low relative to computation costs. For instance, on \dsynthah in Figure~\ref{fig:synthetic}~(a), at $\epsilon=0.05$ \gpu executes in 469~s, and the communication time for the entity partitioning strategy on 128 nodes is only 0.2~s. Therefore, in the case where the dataset cannot be replicated on each GPU, the communication overhead of using a distributed dataset is negligible relative to the total time spent computing the join operation at each node. Furthermore, the entity partitioning strategy does not require any data duplication and achieves a balanced workload.


\subsection{Discussion}
While the scalability analysis is not the focus of this paper, we have shown several insights into scaling \gpu onto multi-GPU systems or distributed-memory nodes in a cluster. We summarize these insights as follows: $(i)$ the fraction of the end-to-end response time associated with overheads such as data transfers and other host-side operations is negligible in high-dimensionality; therefore, multi-GPU systems will not need to contend for limited PCIe bandwidth;  $(ii)$ the entity partitioning strategy we propose achieves excellent load balancing, which is needed to scale \gpu to multi-GPU systems; and, $(iii)$ if the dataset needs to be distributed across compute nodes, the distributed-memory communication cost is low relative to computation. Thus, \gpu is expected to scale well on both multi-GPU systems and clusters with nodes having one or more GPUs.

\section{Conclusions}\label{sec:conclusions}


We summarize the speedup (or slowdown) of \gpu over \ego in Table~\ref{tab:speedup} for two values of $\epsilon$ on each dataset (the lowest and highest $\epsilon$ values examined), and report \gpu configured with the greatest number of optimizations in the respective figures. We find that \gpu outperforms \ego under most experimental scenarios. \ego has excellent dimensionality reordering properties that are able to significantly improve the performance of the algorithm. However, if there is very little variance in the data distribution, then our \gpu approach typically yields substantial performance gains over \ego. Thus, the performance of \gpu may be less sensitive to the data distribution than \ego. On datasets that have been used to compare the performance of self-joins in other works (\colorhist, \layout, and \cooc) \gpu outperforms \ego to a larger degree on the larger values of $\epsilon$ (Table~\ref{tab:speedup}). This occurs because on smaller values of $\epsilon$, the GPU's resources are not saturated, and thus \gpu only yields a minor performance advantage over \ego. 



The self-join is a widely used operation in many data-intensive algorithms. We show that a grid-based index that is suited for the GPU, combined with index dimensionality reduction (indexing $k<n$), reordering the data by the variance in each dimension (\reorder), and distance calculation reduction (\sortidu and \scircuit), outperforms the state-of-the-art self-join, \ego. We also show initial insights into multi-GPU and distributed-memory implementations that are expected to have low communication overheads (over PCIe or between distributed-memory nodes). By utilizing an entity partitioning strategy, we are able to balance the workload across many GPUs, which we expect to enable near-ideal speedups on up to 128 nodes.   

Future work includes developing a performance model to further investigate the effects of our optimizations, and improving the performance of the self-join on small workloads.

\begin{table}[h]
\centering
\caption{Summary of the speedup (or slowdown) of \gpu over \ego on all datasets for the smallest and largest values of $\epsilon$ in the corresponding figures in Section~\ref{sec:expereval}.}\label{tab:speedup}
\begin{tabular}{|c|cc|cc|} \hline
Dataset  &$\epsilon$ & Speedup & $\epsilon$ & Speedup\\ \hline
\colorhist   &0.05&1.13&0.50&5.79\\\hline
\layout  &0.05&1.36&0.50&5.49\\\hline
\cooc    &0.05&3.65&0.20&4.13\\\hline
\susy    &0.01&1.61&0.02&1.07\\\hline
\songs   &0.005&1.53&0.01&0.92\\\hline
\dsynthah&0.03&1.84&0.05&5.46\\\hline
\dsynthai&0.08&8.25&0.10&7.50\\\hline
\dsynthaj&0.16&7.86&0.18&5.63\\\hline

\end{tabular}
\end{table}




\bibliographystyle{ACM-Reference-Format}
\bibliography{bibliography}

\end{document}